\documentclass[one-column]{aastex63}

\usepackage{amsmath}

\newcommand{\swift}{{\it Swift}}

\newcommand{\blue}{\textcolor{blue}}
\newcommand{\src}{Mrk~335}

\usepackage{amsmath}
\usepackage{amsthm}
\usepackage{graphics}
\usepackage{subfigure}
\theoremstyle{definition}
\DeclareMathAlphabet{\mathcal}{OMS}{cmsy}{m}{n} 

\received{March 11, 2020}
\revised{April 10, 2020}
\accepted{April 20, 2021}
\shorttitle{\src}
\shortauthors{Griffiths et al.}
\graphicspath{{./}{figures/}}

\begin{document}

\title{Modelling the Multiwavelength Variability of \src \hspace{0.1em} using Gaussian Processes}

\author{Ryan-Rhys Griffiths}
\affiliation{Department of Physics, University of Cambridge, JJ Thomson Avenue, CB3 0HE Cambridge, UK}

\author{Jiachen Jiang}
\affiliation{Department of Astronomy, Tsinghua University, Beijing 100084, China}
\affiliation{Tsinghua Center for Astrophysics, Tsinghua University, Beijing 100084, China}
\author{Douglas J.K. Buisson}
\affiliation{Department of Astronomy, University of Southampton, Southampton SO17 1BJ, UK}
\author{Dan Wilkins}
\affiliation{Kavli Institute for Particle Astrophysics and Cosmology, Stanford University, 452 Lomita Mall, Stanford, CA 94305, USA}
\author{Luigi C. Gallo}
\affiliation{Department of Astronomy and Physics, Saint Mary's University, 923 Robie Street, Halifax, NS, B3H 3C3, Canada}
\author{Adam Ingram}
\affiliation{Department of Physics, University of Oxford, Keble Road, Oxford OX1 3RH, UK}
\author{Alpha A. Lee}
\affiliation{Department of Physics, University of Cambridge, JJ Thomson Avenue, CB3 0HE Cambridge, UK}
\author{Dirk Grupe}
\affiliation{Department of Physics, Earth Sciences, and Space System Engineering, Morehead State University, Morehead, KY 40514, USA}
\author{Erin Kara}
\affiliation{Kavli Institute for Astrophysics and Space Research, MIT, 77 Massachusetts Avenue, Cambridge, MA 02139, USA}
\author{Michael L. Parker}
\affiliation{Institute of Astronomy, University of Cambridge, Madingley Road, CB3 0HA Cambridge, UK}
\author{William Alston}
\affiliation{European Space Astronomy Centre (ESAC), Science Operations Department, 28692 Villanueva de la Canada, Madrid, Spain}
\author{Anthony Bourached}
\affiliation{UCL Queen Square Institute of Neurology, University College London, Queen Square, London WC1N 3BG, UK}
\author{George Cann}
\affiliation{Department of Space and Climate Physics,  University College London, Mullard Space Science Laboratory, Holmbury St. Mary, Dorking, Surrey RH5 6NT, UK}
\author{Andrew Young}
\affiliation{School of Physics, Tyndall Avenue, University of Bristol, Bristol BS8 1TH, UK}
\author{S. Komossa}
\affiliation{Max-Planck-Institut f\"{u}r Radioastronomie, Auf dem H\"{u}gel 69, 53111 Bonn, Germany}

\begin{abstract}

The optical and UV variability of the majority of AGN may be related to the reprocessing of rapidly-changing X-ray emission from a more compact region near the central black hole. Such a reprocessing model would be characterised by lags between X-ray and optical/UV emission due to differences in light travel time. Observationally however, such lag features have been difficult to detect due to gaps in the lightcurves introduced through factors such as source visibility or limited telescope time. In this work, Gaussian process regression is employed to interpolate the gaps in the Swift X-ray and UV lightcurves of the narrow-line Seyfert 1 galaxy Mrk 335. In a simulation study of five commonly-employed analytic Gaussian process kernels, we conclude that the Matern $\frac{1}{2}$ and rational quadratic kernels yield the most well-specified models for the X-ray and UVW2 bands of Mrk 335. In analysing the structure functions of the Gaussian process lightcurves, we obtain a broken power law with a break point at 125 days in the UVW2 band. In the X-ray band, the structure function of the Gaussian process lightcurve is consistent with a power law in the case of the rational quadratic kernel whilst a broken power law with a break point at 66 days is obtained from the Matern $\frac{1}{2}$ kernel. The subsequent cross-correlation analysis is consistent with previous studies and furthermore, shows tentative evidence for a broad X-ray-UV lag feature of up to 30 days in the lag-frequency spectrum where the significance of the lag depends on the choice of Gaussian process kernel.  

\end{abstract}


\section{Introduction} \label{intro}

Active galactic nuclei (AGN) show strong and variable emission across multiple wavelengths. The UV emission from an AGN is believed to be dominated by thermal emission from an accretion disc close to the central super-massive black hole \citep[SMBH, e.g.][]{pringle81}. {The variability of optical and UV AGN \footnote{AGN with UV and optical luminosity change of more than 1 magnitude such as changing-look AGN, are not discussed in this work cf. \cite{jiang21} for details.} emission is stochastic and described by random Gaussian fluctuations \citep[e.g.][]{welsh11,gezari13,zhu16,sanchez18,smith18,xin20} with the autocorrelation functions of such fluctuations adhering to the `damped random walk' model.} The X-ray emission from an AGN is often found to show faster variability relative to emission at longer wavelengths \citep[e.g.][]{mushotzky93, gaskell03} and originates from a more compact region \citep[e.g.][]{morgan08,chartas17}.

The relationship between the UV and X-ray emission has been well studied. For instance, correlations between the variability in two energy bands has been seen in some individual sources \citep[e.g.][]{shemmer01,buisson17} while others do not show significant evidence for similar correlation \citep[e.g.][]{smith07, 2018_Buisson}. In sources where correlation is found, lags that are related to the light travel time between two emission regions are frequently observed. These lags are often found to be on timescales of days and are longer than those predicted by classical disc theories \citep{shakura73}. Such lag amplitudes indicate a disc of size a few times larger than expected \citep[e.g.][]{edelson00,shappee14,troyer16, buisson17}. {Alternatively, some modified models have been proposed for the underestimation of lags by the classical thin disc model, e.g. disc turbulence \citep[e.g.][]{cai20}, additional varying FUV illumination \citep[e.g.][]{gardner17}, a tilted or inhomogeneous inner disc \citep[e.g.][]{dexter11,starkey17} or an extended coronal region \citep[e.g.][]{kammoun20}.} Much shorter lags, e.g. hundreds of seconds, in agreement with the \citet{shakura73} model have been rarely observed by comparison \citep[e.g. in NGC~4395,][]{mchardy16}.

The Neil Gehrels \swift\ Observatory has been monitoring the X-ray sky in the past decade in tandem with simultaneous pointings in the optical and UV band. In this work, we focus on the X-ray and UVW2 \blue{($\lambda=$212~nm}) lightcurves of the narrow-line Seyfert 1 galaxy \citep[NLS1, e.g.][]{2018_Gallo_New} Mrk~335 obtained by XRT and UVOT, the soft X-ray and UV/optical telescopes on \swift.  Mrk~335 was one of the brightest X-ray sources prior to 2007, before its flux diminished by $10-50\times$ its original brightness \citep{grupe07}.  The X-ray brightness has not recovered since.  During this low X-ray flux period, the UV brightness remains relatively unchanged rendering Mrk~335 X-ray weak \citep{tripathi20}. The behavior has been explained as a possible collapse of the X-ray corona e.g. \citep{2013_Gallo_New, 2015_Gallo_New, 2014_Parker_New} and/or increased absorption in the X-ray emitting region e.g. \citep{grupe12, 2013_Longinotti, 2019_Longinotti, 2019_Parker}.

Mrk~335 has been continuously monitored since 2007 making it one of the best-studied AGN with \swift. Previous studies from the \swift\ monitoring program can be found in \citet{grupe07,grupe12, gallo18, tripathi20, 2020_Komossa}. The X-rays are constantly fluctuating and regularly display large amplitude flaring e.g. \cite{2015_Wilkins}.  The UV are significantly variable, but at a much smaller amplitude than the X-rays.  \cite{gallo18} found tentative evidence for lags of $\approx20$ days based on cross-correlation analyses, suggesting a potential reprocessing mechanism of the more variable X-ray emission in the UV emitter of this source. One challenge faced by the \swift\ monitoring program is that the lightcurves are not continuously sampled and hence standard Fourier techniques cannot be applied. This uneven sampling of the lightcurves is imposed by limited telescope time.

In the context of cross-correlation analysis, methods have been developed to address the problem of unevenly-sampled lightcurves. In \citet{2000_Reynolds}, the method of \citet{1992_Press} is extended to interpolate the lightcurve gaps using a model of the covariance function, or equivalently the power spectrum, of the lightcurve. In \citet{1998_Bond, 2010_Miller, 2013_Zoghbi} a maximum likelihood approach is taken to fit models of the lightcurve power spectra which accounts for the correlation between the lightcurves. In this paper we focus on a relatively new approach to tackle unevenly-sampled lightcurves. 

Gaussian processes confer a Bayesian nonparametric framework to model general time series data \citep{2013_Roberts, 2015_Tobar} and have proven effective in tasks such as periodicity detection \citep{2016_Durrande} and spectral density estimation \citep{2018_Tobar}. More broadly Gaussian processes have recently demonstrated modelling success across a wide range of spatial and temporal application domains including robotics \citep{2011_Deisenroth, 2020_Greeff}, Bayesian optimisation \citep{2015_Shahriari, 2020_Grosnit, 2021_Rivers, 2021_Grosnit} as well as areas of the natural sciences such as molecular machine learning \citep{2021_Nigam, 2020_Griffiths, 2020_flowmo, 2020_Thawani, 2019_Griffiths, 2020_Hase, 2010_Bartok}, genetics \citep{2020_Moss} and materials science \citep{2020_Cheng, 2020_Zhang}. In the context of astrophysics there is a recent trend favouring nonparametric models such as Gaussian processes due to the flexiblity afforded when specifying the underlying data modelling assumptions. Applications have arisen in lightcurve modelling \citep{2021_Luger1, 2021_Luger2, 2021_Luger3}, continuous-time autoregressive moving average (CARMA) processes \cite{2021_Yu}, modelling stellar activity signals in radial velocity data \citep{2015_Rajpaul}, lightcurve detrending \citep{2016_Aigrain}, learning imbalances for variable star classification \citep{2020_Lyon}, inferring stellar rotation periods \citep{2018_Angus}, estimating the dayside temperatures of hot Jupiters \citep{2019_Pass}, exoplanet detection \citep{2017_Jones, 2017_Czekala, 2020_Gordon, 2020_Langellier}, spectral modelling \citep{2012_Gibson, 2018_Nikolov, 2020_Diamond} as well as blazar variability studies \citep{2015_Karamanavis, 2017_Karamanavis, 2020_Covino, 2020_Yang}.

It has recently been demonstrated in lightcurve simulations by \citet{2019_Wilkins} that a Gaussian process framework can compute time lags associated with X-ray reverberation from the accretion disc that are longer and observed at lower frequencies than can be measured by applying standard Fourier transform techniques to the longest available continuous segments. It is for this principal reason that we choose to employ Gaussian processes for our timing analysis. Further desirable facets of Gaussian processes include the fact that, unlike parametric models, they do not make strong assumptions about the shape of the underlying light curve \citep{2012_Wang}. Additionally, we may perform Bayesian model selection at the level of the covariance function or kernel allowing us to quantitatively compare different models of the lightcurve power spectrum. Finally in the cross-correlation analysis, we may make a weaker modelling assumption than \citep{2013_Zoghbi} in treating the X-ray and UV lightcurves as being independent \citep{2019_Wilkins}.

The paper is outlined as follows: In section 2, we provide background on Gaussian processes including discussion of different kernels as well as Bayesian model selection, the criterion used to choose between kernels. In section 3 we describe the procedures used to fit Gaussian processes to the X-ray and UVW2 bands including aspects such as identification of the flux distribution, consideration of measurement noise as well as a simulation study to determine the appropriate kernels. In section 4 we compare the structure functions of the \textsc{gp}-interpolated lightcurves with the observational structure functions from \cite{gallo18}. In section 5 we present a cross-correlation analysis of the X-ray and UVW2 bands using the \textsc{gp}-interpolated lightcurves. Finally, in section 6 we provide concluding remarks about the discrepancy between the observational and \textsc{gp}-derived structure functions as well as the implications of the cross-correlation analysis, namely that the broad lag features suggest an extended emission region of the disc in Mrk 335 during the reverberation process. All code for reproducing the analysis is available at \url{https://github.com/Ryan-Rhys/Mrk_335}.

\section{Gaussian Processes}
\label{gp_intro}

\noindent We may define a Gaussian process (\textsc{gp}) as a collection of random variables, any finite number of which have a joint Gaussian distribution. When the \textsc{gp} is used as a prior over functions, the aforementioned random variables consist of function values $f(t)$ at different points in time $t$. In our setting $f$ represents flux or count rate. The \textsc{gp} is characterised by a mean function

\begin{align}
m(t) = \mathbb{E}[f(t)]
\end{align}

\noindent and a covariance function

\begin{align}
k(t, t') = \mathbb{E}[(f(t) - m(t))(f(t') - m(t))].
\end{align}

\noindent The process is written as follows

\begin{align}
f(t) \sim \mathcal{GP}\big(m(t), k(t, t')\big).
\end{align}

\noindent The mean function is set to the empirical mean of the standardised observational data in the cases we consider. Standardisation, in this case refers to the common practice of subtracting the mean and dividing by the standard deviation of the data when fitting the \textsc{gp} in order to facilitate the identification of appropriate hyperparameters \citep{2008_Murray}. The standardisation is reversed once the fitting procedure is complete in order to obtain predictions on the original scale of the data. $m(t) = 0$ will be assumed henceforth for the sake of the current presentation. The covariance function computes the pairwise covariance between two random variables (function values). In the \textsc{gp} literature, the covariance function is commonly referred to as the kernel and is denoted as

\begin{align}
k(t, t') = \text{cov}\big(f(t), f(t')\big).
\end{align}

\noindent Informally, the kernel is responsible for determining the smoothness of the functions which the \textsc{gp} is capable of fitting. The inductive bias created by the choice of kernel is an important consideration in Gaussian process modelling.

\subsection{Kernels}

\noindent The most widely-known kernel is the squared exponential (SE) or radial basis function (RBF) kernel

\begin{equation}
\label{equation:sqe}
k_{\text{SQE}}(t, t') = \sigma_{f}^2 \exp\bigg(\frac{-|t - t'|^{2}}{2\ell^2}\bigg),
\end{equation}

\noindent where $\sigma_{f}^2$ is the signal amplitude hyperparameter (vertical lengthscale) and $\ell$ is the (horizontal) lengthscale hyperparameter. For such hyperparameters, we will adopt the notation of $\theta$ to represent the set of kernel hyperparameters. It has been argued by \cite{2012_Stein} that the smoothness assumptions of the squared exponential kernel are unrealistic for many physical processes. As such, kernels such as the Matern

\begin{equation}
k_{\text{Matern}}(t, t') = \frac{2^{1-\nu}}{\Gamma(\nu)}\bigg(\frac{\sqrt{2\nu}-|t - t'|}{\ell}\bigg)^\nu K_{\nu}\bigg(\frac{\sqrt{2\nu}-|t - t'|}{\ell}\bigg),
\end{equation}

\noindent are more commonly seen in the machine learning literature. Here $K_{\nu}$ is a modified Bessel function of the second kind, $\Gamma$ is the gamma function and $\nu$ is a non-negative parameter of the kernel which is typically taken to be either $\frac{3}{2}$ or $\frac{5}{2}$ \citep{2006_Rasmussen}. The lengthscale hyperparameter $\ell$ can be thought of loosely as a decay coefficient for the covariance between inputs as they become increasingly far apart in the input space; the further apart the inputs are, the less correlated they will be. The final kernel used in this work is the rational quadratic (RQ) kernel

\begin{equation}
k_{RQ}(t, t') = \bigg(1 + \frac{|t - t'|^{2}}{2\alpha\ell^2}\bigg)^{-\alpha}.
\end{equation}

\noindent where $\alpha, \ell > 0$. The rational quadratic kernel can be viewed as a scale mixture of squared exponential kernels with different characteristic lengthscales. All kernels used in this work are stationary kernels and as such it should be stated that this reflects a modelling assumption that the underlying time series is stationary. The extension of the current work to include non-stationary kernels will be discussed in \autoref{conc}.

\subsection{Prediction with Gaussian Processes}

To illustrate the homoscedastic (constant noise) \textsc{gp} predictive model we use X-ray timing as an example. We wish to model the count rate $f(t)$. We place a Gaussian process prior over $f$,

\begin{align}
p(\mathbf{f}(\mathbf{t})| \theta) = \mathcal{GP}\big(0, K(\mathbf{t}, \mathbf{t})\big)
\end{align}

\noindent where $\mathbf{f}$ denotes the vector of function values evaluated at the set of times $\{t_{i}\}_{i=1}^N, N \in \mathbb{N}$. $K(\mathbf{t}, \mathbf{t})$ is a kernel matrix where entries are computed by the kernel function as $[K]_{ij} = k(t_i, t_j)$. $\theta$ represents the set of kernel hyperparameters. The Gaussian process prior is written as

\begin{align}
p(\mathbf{y}(\mathbf{t})|\theta) = \mathcal{GP}\big(0, K(\mathbf{t}, \mathbf{t}) + I \sigma_{y}^2\big),
\end{align}

\noindent where $I\sigma_{y}^2$ represents the variance of iid Gaussian noise on the observations $\mathbf{y}$. The applicability of such a noise model, also known as a Gaussian likelihood, will be discussed further in \autoref{noise}. Once we have observed some data $\mathbf{y}$, the joint distribution over the observed data $\mathbf{y}$ and the predicted function values $\mathbf{f}*$ at test locations $\textbf{t}*$ may be written as

\begin{align}
    \begin{bmatrix} 
        \mathbf{y} \\
        \mathbf{f_*} \\
    \end{bmatrix}
    \sim
    \mathcal{N}
    \bigg(0,
    \begin{bmatrix}
        K(\mathbf{t}, \mathbf{t}) + \sigma_{y}^2 I & K(\mathbf{t}, \mathbf{t_*})\: \\
        K(\mathbf{t_*}, \mathbf{t})\phantom{+ \: \: \sigma{y}^2} & K(\mathbf{t_*}, \mathbf{t_*})
    \end{bmatrix}
    \bigg),
\end{align}

\noindent where $\mathcal{N}$ is the multivariate Gaussian probability density function. The joint prior may be conditioned on the observations through

\begin{align}
p(\mathbf{f_*}| \mathbf{y}) = \frac{p(\mathbf{f_*}, \mathbf{y})}{p(\mathbf{y})}
\end{align}

\noindent which enforces that the joint prior agrees with the observed target values $\mathbf{y}$. The predictive distribution is thus given as 

\begin{align}
p(\mathbf{f_*}| \mathbf{t}, \mathbf{y}, \mathbf{t_*}) = \mathcal{N}\big(\mathbf{\bar{f}_*}, \text{cov}(\mathbf{f_*})\big)
\end{align}

\noindent with the predictive mean at test locations $\mathbf{t_*}$ being

\begin{align}
\mathbf{\bar{f_*}} = K(\mathbf{t_*}, \mathbf{t})[K(\mathbf{t}, \mathbf{t}) + \sigma_{y}^2 I]^{-1} \mathbf{y}
\end{align}

\noindent and the predictive uncertainty being

\begin{align}
\text{cov}(\mathbf{f_*}) = K(\mathbf{t_*}, \mathbf{t_*}) - K(\mathbf{t_*}, \mathbf{t})[K(\mathbf{t}, \mathbf{t}) + \sigma_{y}^2 I]^{-1} K(\mathbf{t}, \mathbf{t_*}).
\end{align}

Analysing the form of this expression one may notice that the first term $K(\mathbf{t_*}, \mathbf{t_*})$ in the expression for the predictive uncertainty $\text{cov}(\mathbf{f}*)$ may be viewed as the prior uncertainty and the second term $K(\mathbf{t_*}, \mathbf{t})[K(\mathbf{t}, \mathbf{t}) + \sigma_{y}^2 I]^{-1} K(\mathbf{t}, \mathbf{t_*})$  can be thought of as a subtractive factor that accounts for the reduction in uncertainty when observing the data points $\mathbf{y}$.

\subsection{Bayesian Model Selection}

One desirable property of Gaussian processes and Bayesian models in general is the ability to carry out hierarchical modelling \citep{1992_MacKay_Hierarchical, 2019_der_Wilk}. The three tiers of the modelling hierarchy are:

\begin{enumerate}
    \item Model Parameters
    \item Model Hyperparameters
    \item Model Structures
\end{enumerate}

\noindent In the case of the nonparametric Gaussian process framework, parameters do not have the same meaning as in parametric Bayesian models and are instead obtained from the posterior distribution over functions. Hyperparameters are typically parameters of the kernel function such as signal amplitudes and lengthscales. An important entity for hyperparameter optimisation in Gaussian processes is the log marginal likelihood or evidence \citep{1992_MacKay}

\begin{align}
\label{equation: log_lik}
\log p(\mathbf{y}| \mathbf{t}, \theta) =&  \underbrace{-\frac{1}{2} \mathbf{y}^{\top}(K_{\theta}(\mathbf{t}, \mathbf{t}) + \sigma_{y}^2I)^{-1} \mathbf{y}}_\text{encourages fit with data} \\ 
&\underbrace{-\frac{1}{2} \log | K_{\theta}(\mathbf{t}, \mathbf{t}) + \sigma_{y}^2 I |}_\text{controls model capacity} -\frac{N}{2} \log(2\pi) \nonumber.
\end{align}

Where $N$ is the number of observations and the subscript notation on the kernel matrix $K_{\theta}(\mathbf{t}, \mathbf{t})$ is chosen to indicate the dependence on the set of hyperparameters $\theta$. The two terms in the expression for the marginal likelihood embody Occam's Razor \citep{2001_Rasmussen} in their preference for selecting models of intermediate capacity. The first term in \autoref{equation: log_lik} acts as a term that penalises functions that do not fit the data well whereas the second term acts like a regulariser, disfavouring overly complex models. In this work kernel hyperparameters are chosen to optimise the marginal likelihood. At the level of model structures, the fit achieved by different kernels can be quantitatively assessed by comparing the values of the optimised log marginal likelihood objective.

\newpage

\section{Gaussian Process Modelling of Mrk 335} 

\noindent In this paper we consider the Swift X-ray and UVW2 lightcurves in time bins of one day. We refer the reader to \cite{gallo18} for details of the data reduction processes. The observational measurements used in this work run from $54327-58626$ modified Julian days and comprise $509$ data points for the X-ray band and $498$ data points for the UVW2 band. We consider the latest UVOT sensitivity calibration file (`swusenscorr20041120v006.fits') to account for the sensitivity loss with time in the UVW2 band\footnote{The most up-to-date calibration files may be found at \href{https://heasarc.gsfc.nasa.gov/docs/heasarc/caldb/swift}{https://heasarc.gsfc.nasa.gov/docs/heasarc/caldb/swift}. We consider only UVW2 data collected by UVOT because the UVW2 filter was most frequently-used in the archival observations.}.

\subsection{Identifying the Flux Distribution}

\begin{figure*}[]
\centering
\subfigure[X-Ray Log Count Rates]{\label{fig:4}\includegraphics[width=0.49\textwidth]{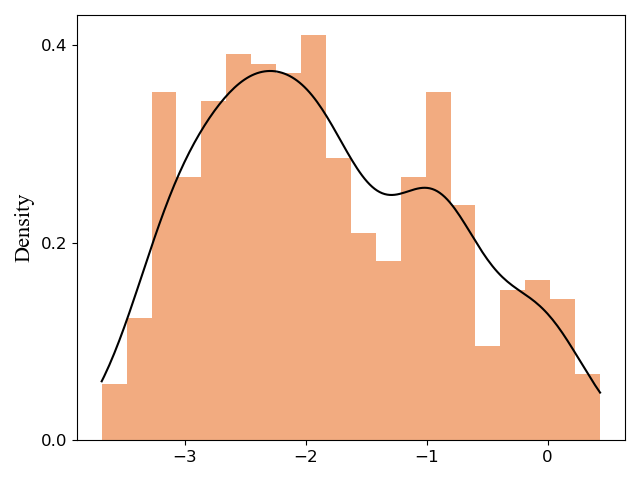}}
\subfigure[UVW2 Flux]{\label{fig:3}\includegraphics[width=0.49\textwidth]{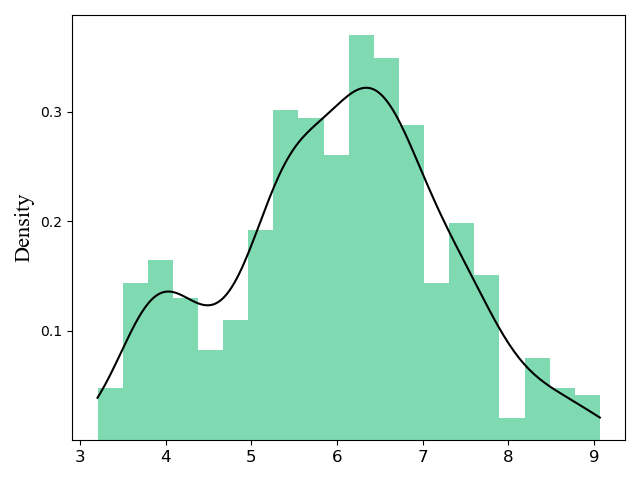}}
\caption{Histograms of the observed Swift X-ray log count rate and UVW2 flux overlaid with Gaussian kernel density estimates. The raw UVW2 flux values have been scaled by $1e^{14}$.}
\label{Histograms}
\end{figure*}

In order to assess the applicability of Gaussian processes in modelling the flux distribution of the X-ray and UVW2 bands of Mrk 335, we perform a series of graphical distribution tests to determine the sample distribution. The histograms of the log count rates for the X-ray, and flux for the UV bands, of Mrk 335 are shown in \autoref{Histograms}. The histograms show that the distribution of the UVW2 flux is approximately Gaussian-distributed whereas the X-ray count rate distribution appears to be log-Gaussian distributed in line with the general observation of \cite{2005_Uttley} that fluxes from accreting black holes tend to follow log-Gaussian distributions. We provide further graphical distribution tests based on probability-probability (PP) plots and empirical cumulative distribution functions (ECDFs) in \autoref{dist_tests}. 

Furthermore, following \cite{2019_Wilkins} we perform a Kolmogorov-Smirnov test for goodness-of-fit where the null hypothesis is that the sample was drawn from a Gaussian distribution. For the UVW2 flux values we obtain a p-value of $0.164$. We obtain a p-value of $1.017e^{-20}$ for the raw X-ray count rates and a p-value of $0.028$ for the log-transformed X-ray count rates. As such, we cannot reject the null hypothesis that either UVW2 flux or log-transformed X-ray count rates are drawn from a Gaussian distribution at the $1\%$ level of significance. We may however reject the null hypothesis in the case of the raw X-ray count rates, providing evidence that the raw X-ray count rates should be log-transformed in order to be well-modelled by a Gaussian distribution. As such, we log transform the raw X-ray count rates and leave the UVW2 flux values unchanged.

\newpage

\subsection{Noise}
\label{noise}

As noted by \cite{2019_Wilkins} fitting a Gaussian process to the logarithm of the count rate is appropriate only in the limit of a large signal-to-noise ratio. In the case of Mrk 335, the Poisson (shot) noise intrinsic to the photon detectors used to obtain the flux measurements is over an order of magnitude smaller than the flux measurement itself. As such the choice of the log-Gaussian process would appear to be justified. 

\subsection{Simulations}
\label{sim_section}

\begin{figure*}[h]
    \centering
    \includegraphics[width=0.49\textwidth]{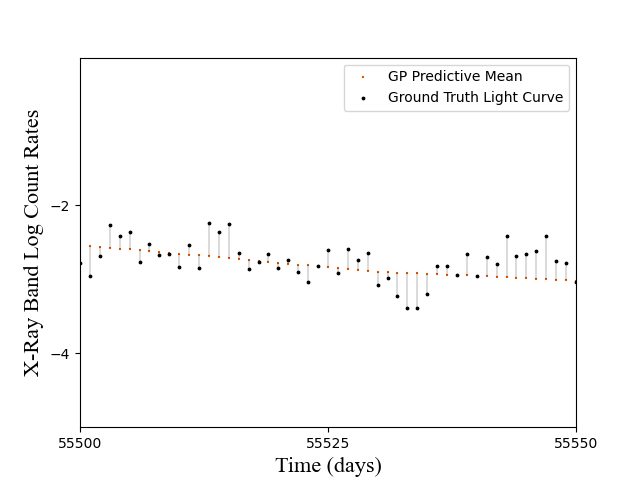}
    \caption{Residual plot. The normalised RSS metric is the sum of squared residuals divided by the total number of discretised points (4390) comprising the simulated lightcurve. A residual in this case represents the difference between the Gaussian process predictive mean and the ground truth value of the simulated lightcurve.}
    \label{rss_figure}
\end{figure*}

We undertake a simulation study in order to quantitatively assess the abilities of different kernels to interpolate gapped simulated lightcurves. Observational power spectral densities (PSDs) of AGN are well-described by (broken) power laws \citep{2004_Mchardy}. As such, our simulations employ a power law PSD with index fit to the observational data. Our goals with the study are twofold: Firstly, although we cannot be sure of the true PSD for the observational data, we hope that the simulations may afford a good proxy for identifying performant kernels based on the fact that AGN typically exhibit power law-like PSDs and secondly, we wish to test whether a kernel's ability to reconstruct the full simulated lightcurve correlates with its marginal likelihood value for the gapped data on which it is trained. If there is a correlation, we may use the marginal likelihood as a metric for identifying the appropriate kernel on the observational data.

One thousand simulated light curves with gaps are generated for the \src \: X-ray and UV bands using the method of \cite{1987_Davies}, first applied in astrophysics by \cite{1995_Timmer}. For each lightcurve we have access to the ground truth functional form of the lightcurve before the introduction of gaps. Computationally, the ground truth lightcurve is evaluated on a fine, discrete grid of $4390$ time points whereas the gapped lightcurves are evaluated on a coarser, unevenly-spaced grid of $498$ time points for the UV simulations and $509$ time points for the X-ray simulations in line with the number of observational data points. We then quantify how well each \textsc{gp} kernel performs in recovering the ground truth lightcurve by measuring the normalised residual sum of squared errors,

\begin{align}
    \text{RSS} = \frac{1}{N} \sum_{i=1}^{N} (f(t_i) - y_i)^2
\end{align}

where $f(t_i)$ is the Gaussian process prediction at grid point $t_i$ and $y_i$ is the true simulated count rate value. The RSS values are averaged over the one thousand simulated lightcurves. We provide an illustration of the RSS metric in \autoref{rss_figure}. In addition, we compute the averaged negative log marginal likelihood (NLML) for each kernel:

\begin{align}
    \text{NLML} =  \frac{1}{2} \mathbf{y}^{\top}(K_{\theta}(\mathbf{t}, \mathbf{t}) + \sigma_{y}^2I)^{-1} + \mathbf{y} \frac{1}{2} \log | K_{\theta}(\mathbf{t}, \mathbf{t}) + \sigma_{y}^2 I | + \frac{N}{2} \log(2\pi).
\end{align}

\noindent The NLML in this case is the negative of the quantity given in \autoref{equation: log_lik}. Kernel hyperparameters were selected via optimisation of the NLML using the scipy optimiser of GPflow \citep{GPflow}. The jitter level was fixed at 0.001, a small positive number to ensure numerical stability. The output values (flux or the logarithm of the count rate) were standardised according to their empirical mean and standard deviation. We use a constant mean function set to the empirical mean of the data following standardisation as discussed in \autoref{gp_intro}.

We report the results of this simulation study in \autoref{sim_study}. The NLML values show correlation with RSS, thus providing evidence that NLML is an appropriate metric for determining the Gaussian process kernel for the real observational data (for which the ground truth lightcurve is of course not available). A paired t-test was conducted to determine whether the RSS results were significant in terms of identifying the best kernel. For the X-ray simulations, a t-statistic of $9$ was obtained corresponding to a two-sided p-value of $5^{-20}$. For the UVW2 simulations, a t-statistic of $-22$ was obtained corresponding to a two-sided p-value of $9^{-85}$. As such, the null hypothesis that the performance discrepancy between kernels on the RSS metric is due to chance variation across $1000$ simulations, may be rejected at the $1\%$ level of significance. We offer further rationalisation in \autoref{kern_rat} for why the top two performing kernels in the simulation study are the Matern $\frac{1}{2}$ and rational quadratic kernels.

\begin{table}[h]
\caption{Performance comparison of kernels based on the NLML on the simulated gapped X-ray and UV lightcurves and normalised residual sum of squared errors (RSS) on the ground truth simulated lightcurves. The mean NLML and RSS across 1000 simulations are reported with the standard error. UVW2 RSS values have an exponent of $-30$. Bold values indicate best performance.}
\label{sim_study}
\begin{center}
\begin{tabular}{l|ll}
\hline
Kernel & NLML & RSS \\ \hline
\multicolumn{1}{c|}{\underline{\textbf{X-Ray}}} & & \\[5 pt]
Matern$\frac{1}{2}$ & $\textbf{180.2} \pm \: \textbf{3.8}$ & $0.121 \pm \: 0.002$  \\
Matern$\frac{3}{2}$ & $420.7 \pm \: 3.3$ & $0.309 \pm \: 0.003$ \\
Matern$\frac{5}{2}$ & $523.5 \pm \: 2.9$ & $0.374 \pm \: 0.003$ \\
Rational Quadratic & $\textbf{184.2} \pm \: \textbf{3.6}$ & $\textbf{0.117} \pm \: \textbf{0.002}$ \\
Squared Exponential & $632.1 \pm \: 1.5$ & $0.554 \pm \: 0.004$ \\ \hline
 \multicolumn{1}{c|}{\underline{\textbf{UVW2}}} & & \\[5 pt]
Matern$\frac{1}{2}$ & $\textbf{-399.0} \pm \: \textbf{5.2}$ & $\textbf{2.9} \pm \: \textbf{0.08}$ \\
Matern$\frac{3}{2}$ & $-298.3 \pm \: 6.0$ & $7.9 \pm \: 0.25$ \\
Matern$\frac{5}{2}$ & $-219.6 \pm \: 6.5$ & $17.0 \pm \: 0.41$ \\
Rational Quadratic & $-349.2 \pm \: 5.4$ & $3.4 \pm \: 0.09$ \\
Squared Exponential & $-65.0 \pm \: 7.4$ & $32.8 \pm \: 0.55$ \\ \hline
\end{tabular}
\end{center}
\end{table}


\subsection{Gaussian Process Fits}

\begin{figure*}[]
\centering
\subfigure[UV Band | Matern $\frac{1}{2}$ | Mean.]{\label{fig:4pp1}\includegraphics[width=0.49\textwidth]{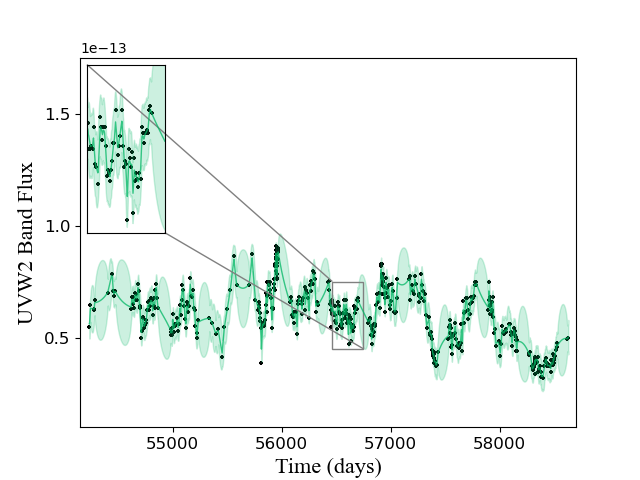}}
\subfigure[UV Band | Matern $\frac{1}{2}$ | Sample.]{\label{fig:4pp2}\includegraphics[width=0.49\textwidth]{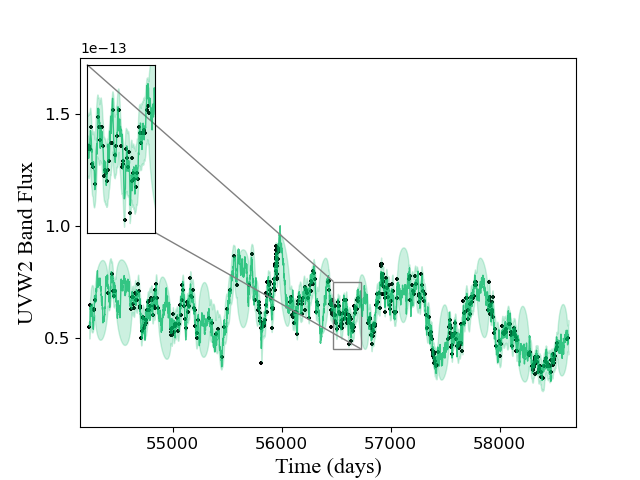}}
\subfigure[UV Band | Rational Quadratic | Mean.]{\label{fig:4pp3}\includegraphics[width=0.49\textwidth]{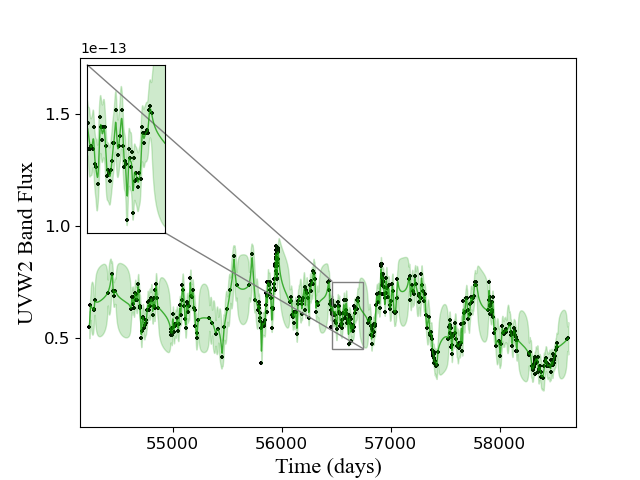}}
\subfigure[UV Band | Rational Quadratic | Sample]{\label{fig:4pp4}\includegraphics[width=0.49\textwidth]{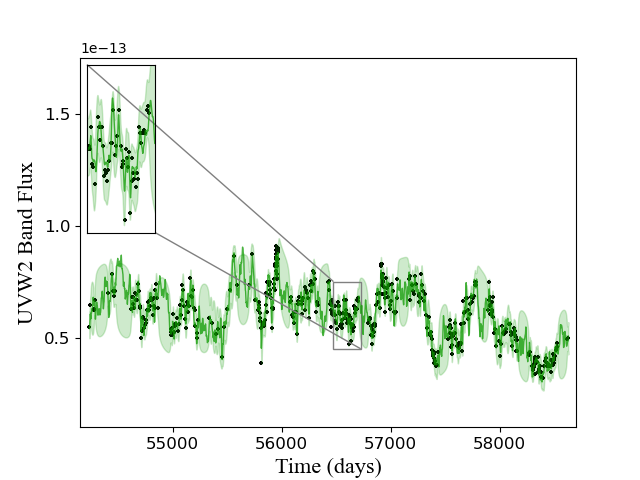}}  
\caption{\textsc{gp} lightcurves for the UVW2 band. The shaded regions denote the \textsc{gp} 95\% confidence interval.} We show both the \textsc{gp} mean and a sample from the \textsc{gp} posterior in separate plots. The insets are included to highlight the variability of the fit.
\label{GP_uv_fits}
\end{figure*}

\begin{figure*}[]
\centering
\subfigure[X-ray Band | Matern $\frac{1}{2}$ | Mean.]{\label{fig:4pp}\includegraphics[width=0.49\textwidth]{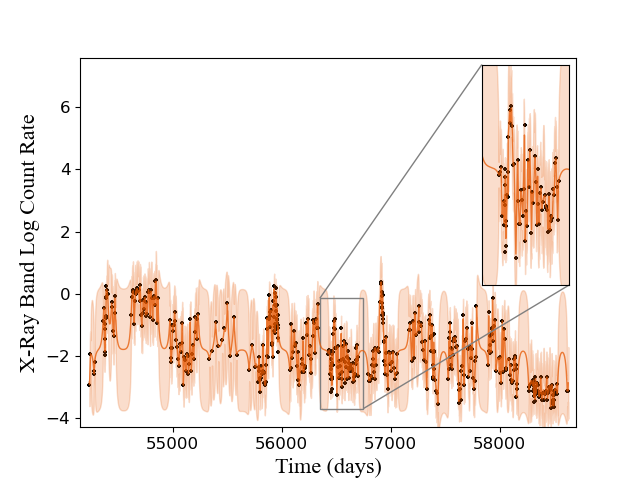}}
\subfigure[X-ray Band | Matern $\frac{1}{2}$ | Sample.]{\label{fig:4pp}\includegraphics[width=0.49\textwidth]{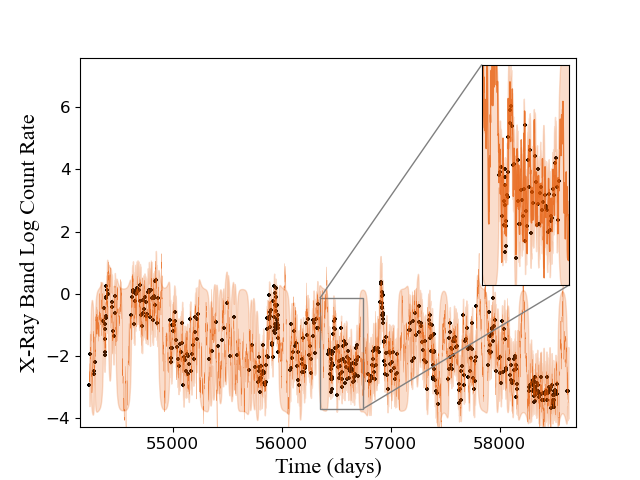}}
\subfigure[X-ray Band | Rational Quadratic | Mean.]{\label{fig:4pp}\includegraphics[width=0.49\textwidth]{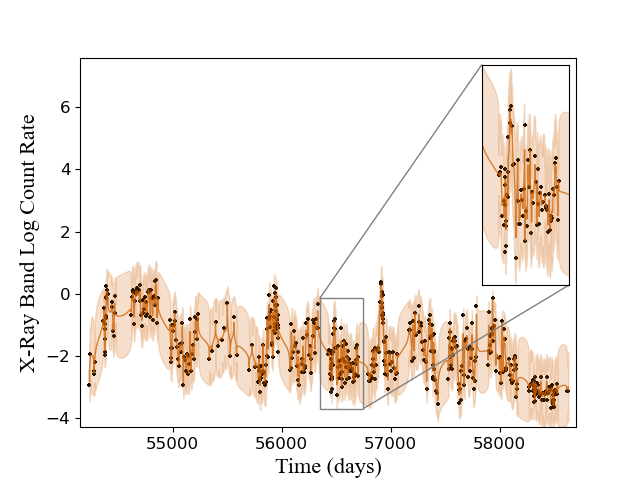}}
\subfigure[X-ray Band | Rational Quadratic | Sample]{\label{fig:4pp}\includegraphics[width=0.49\textwidth]{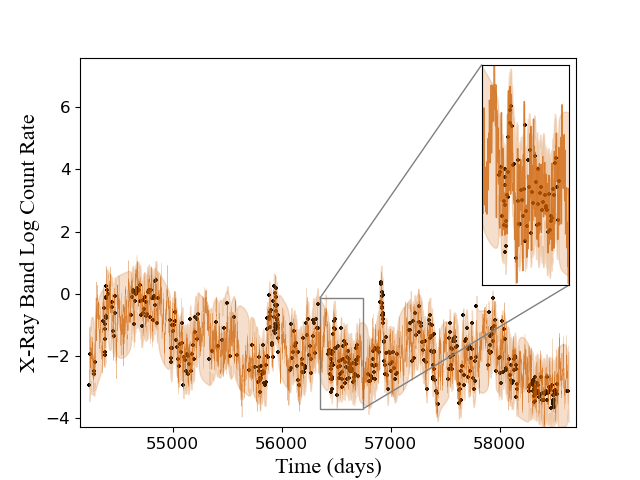}}  
\caption{\textsc{gp} lightcurves for the X-ray band. The shaded regions denote the \textsc{gp} 95\% confidence interval.} We show both the \textsc{gp} mean and a sample from the \textsc{gp} posterior in separate plots. The insets are included to highlight the variability of the fit.
\label{GP_xray_fits}
\end{figure*}

The fits to the observational data for the UVW2 and X-ray bands are shown in \autoref{GP_uv_fits} and \autoref{GP_xray_fits} respectively. In an analogous fashion to the simulation experiments we evaluate five stationary kernels: Matern $\frac{1}{2}$, Matern $\frac{3}{2}$, Matern $\frac{5}{2}$, rational quadratic and squared exponential. We choose to display the two kernels, rational quadratic and Matern $\frac{1}{2}$ which performed best in the simulation study in their abilities to model power law-like PSDs. These kernels also have the most favourable values under the NLML metric for the observational data. We again use a constant mean function set to the empirical mean of the data following standardisation. We optimise all kernel hyperparameters under the marginal likelihood save for the noise level which we fix to a constant value in the standardised space. This constant noise value is computed by dividing the mean output value in the standardised space by the mean signal-to-noise ratio in the original space.

\newpage

\section{Structure Function Analysis}

Ideally we would like to examine the PSD of the Gaussian process fits to the observational data. The PSD characterises the distribution of power over frequencies of a given emission band and properties of the PSD can be linked to underlying physical processes in the accretion disk. Computation of the PSD, while possible, can be complicated by the uneven sampling of the observational data, leading previous studies to instead perform a structure function analysis on the \src \: data \citep{gallo18}. While it is possible to extract the PSD from the learned kernel \citep{2019_Wilkins}, in this work we choose to perform a structure function analysis of the \textsc{gp} lightcurves in order to compare directly against the results of \cite{gallo18}. We follow the method described in \cite{1985_Simonetti, 1992_Hughes, 1996_di_Clemente, 2001_Collier, gallo18}. The binned structure function is defined as:

\begin{align}
    \text{SF}(\tau) = \frac{1}{N(\tau)} \sum_{i} \: [f(t_i) - f(t_i + \tau)]^2
\end{align}

where $\tau = t_j - t_i$ is the distance between pairs of points $i$ and $j$ such that $t_j > t_i$. The structure function is binned according to $\tau$ where the centres of each bin are given by $\tau_i = (i - \frac{1}{2})\delta$. $\delta$ in this instance is the structure function resolution. We use the same $\delta$ as in \cite{gallo18}, namely 5.3 days for the structure function computation over both the X-ray and UVW2 bands. $f(t_i)$ gives the count rate value at time point $t_i$ and $N(\tau)$ is the number of structure function pairs in each bin $i$ with centre $\tau_i$. Accounting for measurement noise by subtracting twice the mean noise variance from each structure function bin, as performed in \cite{gallo18} was found to have negligible effect on the \textsc{gp} structure functions and so we ignore it. As in \cite{gallo18}, we normalise the structure function values by the global lightcurve variance. 

\begin{figure*}[]
\centering
\subfigure[Matern $\frac{1}{2}$ UVW2]{\label{fig:4pp}\includegraphics[width=0.49\textwidth]{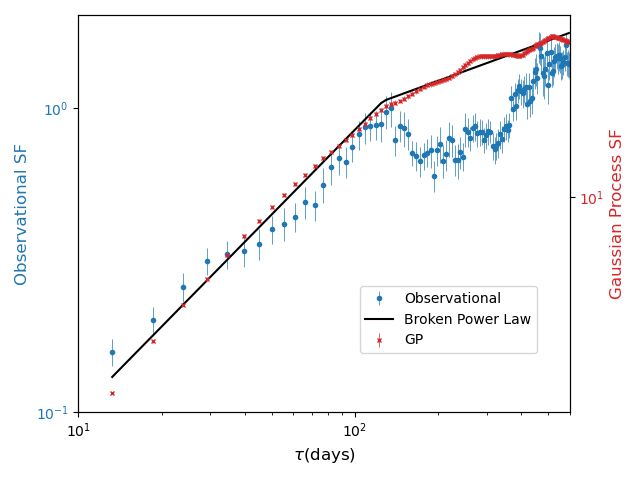}}
\subfigure[Rational Quadratic UVW2]{\label{fig:4pp}\includegraphics[width=0.49\textwidth]{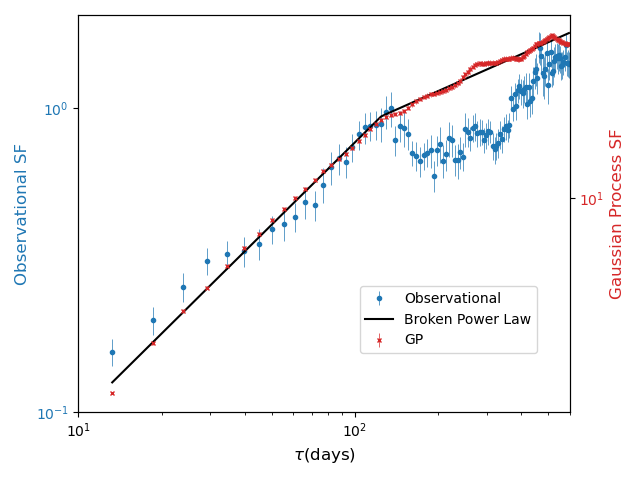}}  
\subfigure[Matern $\frac{1}{2}$ X-ray]{\label{fig:4pp}\includegraphics[width=0.49\textwidth]{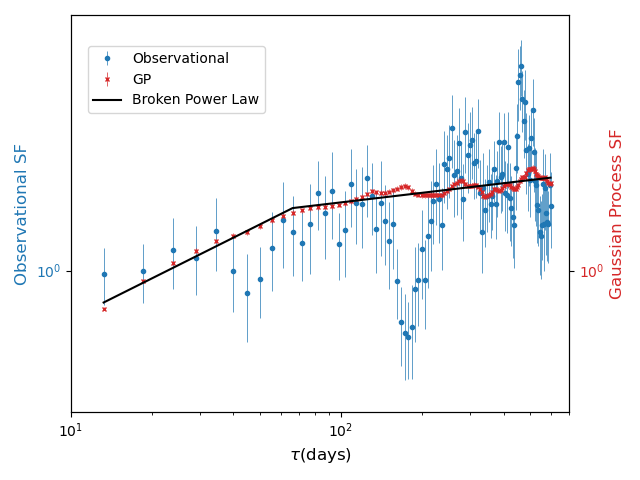}}
\subfigure[Rational Quadratic X-ray]{\label{fig:4pp}\includegraphics[width=0.49\textwidth]{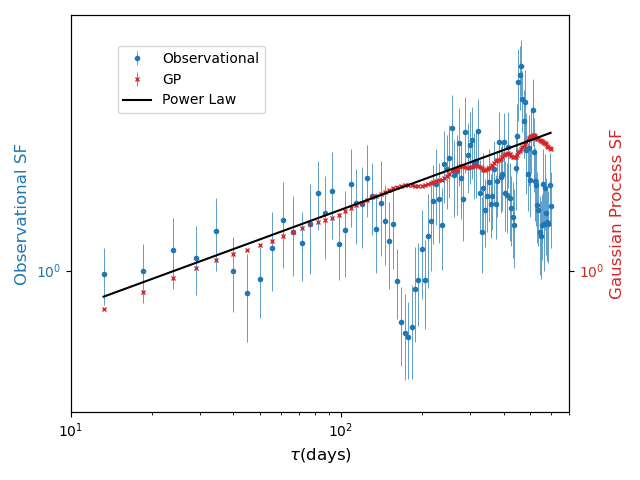}}
\caption{Comparison of observational and \textsc{gp} structure functions. The \textsc{gp} structure functions are consistent with those calculated from the observational data in the non-noise dominated regions. The dip at ca. $200$ days in the observational X-ray structure function is potentially a sampling artefact as demonstrated by simulation in \autoref{sf_sims}.}
\label{GP Structure Functions}
\end{figure*}

The \textsc{gp} structure functions for the interpolated lightcurves are shown in \autoref{GP Structure Functions}. The $1\sigma$ \textsc{gp} error bars are obtained by computing the structure function over 50 samples from the Gaussian process posterior. Each sample gives rise to highly similar structure functions and so the errors are not visible on the plot. The structure functions computed from the observational data, 509 and 498 data points for the X-ray and UV bands of \src \: respectively, are included for reference. In contrast to the \textsc{gp} structure function errorbars, in the case of the observational data the error bars are computed as $\frac{\sigma_i}{\sqrt(\frac{N_i}{2})}$ where $\sigma_i$ is the noise standard deviation in bin $i$ and $N_i$ is the number of pairs in bin $i$.\\ 

The Gaussian process structure functions are compared against the observational structure functions in \autoref{GP Structure Functions}. In addition, we plot broken power law fits to the \textsc{gp} structure functions, the parameters of which are given in \autoref{params}. In the UVW2 band, both Gaussian process kernels yield structure functions possessing a consistent break point at ca. 125 days. In the X-ray band the Matern $\frac{1}{2}$ kernel yields a break point at 66 days whereas the rational quadratic kernel fit yields an unbroken power law. Given the discrepancy between \textsc{gp} kernels, we do not find definite evidence for a break in the X-ray power law.

\begin{figure*}[]
\centering
\subfigure[Observational]{\label{fig:sf_sim}\includegraphics[width=0.49\textwidth]{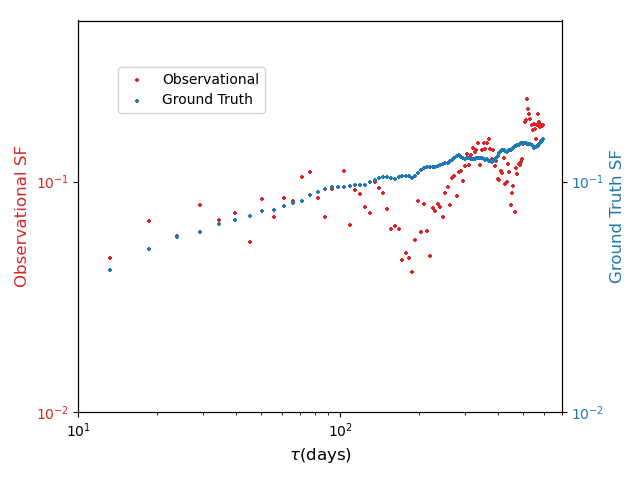}}
\subfigure[\textsc{gp} | Matern $\frac{1}{2}$]{\label{fig:sf_sim2}\includegraphics[width=0.49\textwidth]{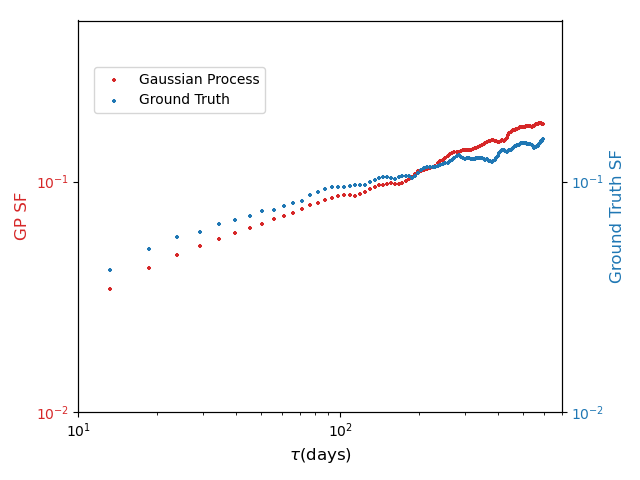}}
\caption{Structure function simulations. Pseudo-observational lightcurves are produced by introducing gaps into the simulated lightcurves. The structure function for the gapped lightcurve is shown in red in (a) whereas the structure function of the \textsc{gp} interpolation is shown in red in (b). Both structure functions are compared against the ground truth structure function obtained from the full simulated ground truth lightcurve (no gaps). The dips at $\tau = 200 \: \text{days}$ and $\tau = 400 \: \text{days}$ in the structure function derived from the gapped observational simulation in 6(a) are artefacts of the uneven sampling.}
\label{sf_sims}
\end{figure*}

Of particular interest is whether the dip in the X-ray structure function is an expected feature of the latent lightcurve or a measurement artefact arising from uneven sampling. In order to assess the potential for the dip to be a sampling artefact, we perform simulations using the Timmer and Konig algorithm from \autoref{sim_section}. In this case we compute structure functions of gapped lightcurves and compare against structure functions derived from the ground truth lightcurves with no gaps. One representative simulation is depicted in \autoref{sf_sims}. In this instance a similar dip to that found in the observational data is observed in the X-ray band simulation. This highlights the possiblity that the dip seen in the observational X-ray structure function is a sampling artefact arising from gaps in the lightcurve.

\begin{table}[]
\caption{Parameters for the broken power law fits to the \textsc{gp} structure functions. $\alpha_1$ and $\alpha_2$ are the indices for the power law before and after the break point $\tau_{char}$. The break point $\tau_{char}$ is reported in days. Errors are computed using 200 bootstrap samples of the data points corresponding to the GP structure functions. The X-ray rational quadratic structure function is fit using a power law and as such only has a single index as a parameter. The astropy library \citep{astropy_1, astropy_2} is used to compute the (broken) power law fits using the simplex algorithm and least squares statistic for optimisation with the \textsc{gp} structure function uncertainties used as weights in the fitting.}
\begin{center}
\begin{tabular}{lllll}
\hline
Waveband & Kernel & $\alpha_1$ & $\alpha_2$ & $\tau_{char}$ \\ \hline
UVW2 & Matern $\frac{1}{2}$ & $-0.72 \pm \: 0.03$ & $-0.26 \pm \: 0.01$ & $127 \pm \: 8$ \\
UVW2 & Rational Quadratic & $-0.62 \pm \: 0.01$ & $-0.28 \pm \: 0.01$ & $125 \pm \: 5$ \\
X-ray & Matern $\frac{1}{2}$ & $-0.29 \pm \: 0.03$ & $-0.07 \pm \: 0.004$ & $66 \pm \: 8$ \\
X-ray & Rational Quadratic & $-0.21 \pm \: 0.002$ & N/A & N/A \\ \hline
\end{tabular}
\end{center}
\label{params}
\end{table}

\newpage

\section{Lag and Coherence}

\begin{figure*}
    \centering
    \includegraphics{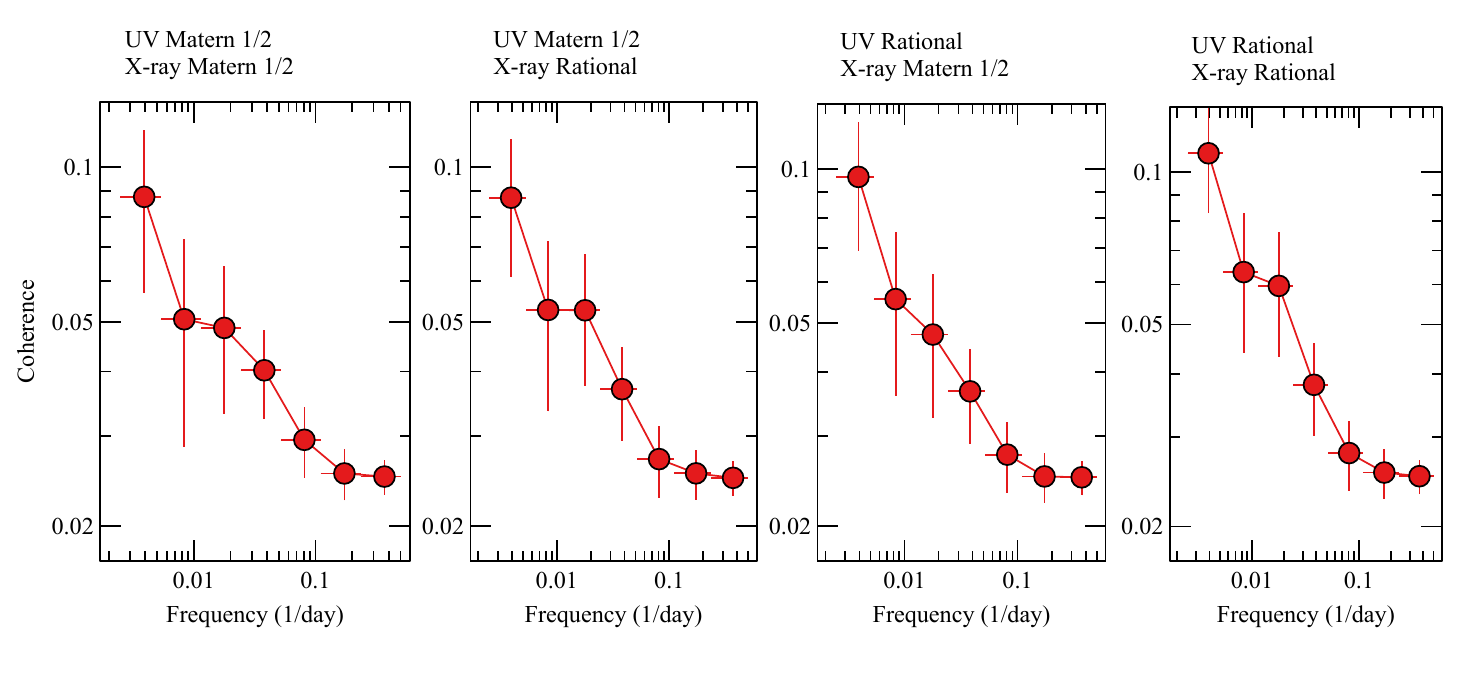}
    \includegraphics{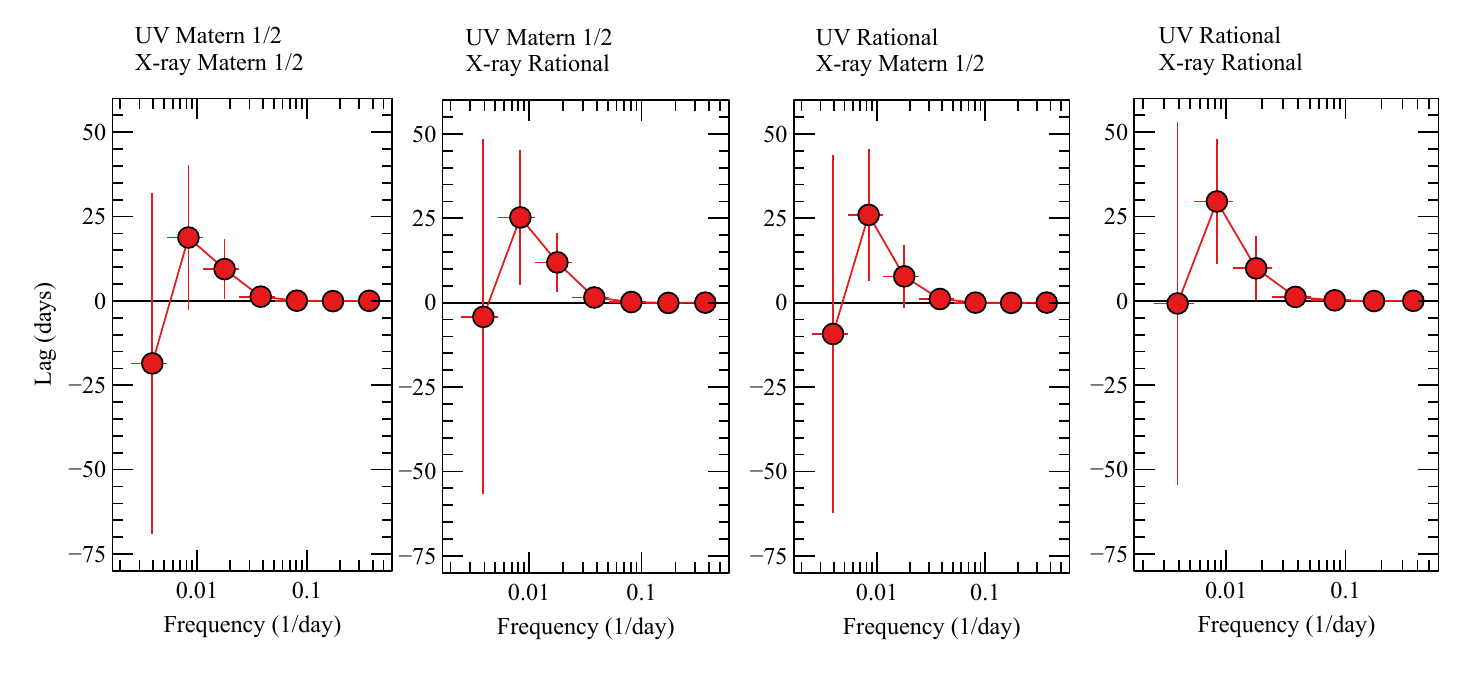}
    \caption{The coherence and lag spectra for Mrk~335. They are calculated by using 1000 pairs of \textsc{gp} lightcurve samples fit to the observed lightcurves. The error bars are the standard errors of the corresponding measurements for the 1000 samples. Different panels are for different kernels. Positive lags imply that the X-ray band leads the UVW2 band.}
    \label{pic_coh_lag}
\end{figure*}

In this section, we calculate the coherence between the UVW2 and X-ray emission from \src \: in search of evidence of lag features in the Fourier frequency domain. The coherence and lag spectra are estimated from one thousand pairs of UVW2 and X-ray \textsc{gp} lightcurve samples drawn from the Gaussian process posterior for each kernel. The lags are defined as the phase lags divided by the corresponding Fourier frequency. {A similar approach has been used in other disciplines \citep[e.g.][]{fabian09, kara13}.} We consider both Matern $\frac{1}{2}$ and Rational Quadratic kernels. The results are shown in \autoref{pic_coh_lag}. Positive lags imply that the X-ray variability leads the UVW2 variability. The error bars in the figure are the standard errors of the corresponding measurements for the one thousand samples.

The coherence between the UVW2 and X-ray emission decreases with frequency, suggesting more coherent variability at lower frequency. Positive lag features are shown at the low frequencies in the range $f=$ 0.005--0.025 d$^{-1}$. The absolute value of the lag at $f=0.0039\pm0.0014$\,d$^{-1}$ is estimated to be $19 \pm 22$ days for the Matern $\frac{1}{2}$ kernel applied to both lightcurves and $29 \pm 19$ days for the rational quadratic kernel, however both measurements are consistent with zero lag in the $2\sigma$ uncertainty range.

Tentative evidence of a shorter time lag at a higher frequency of $f=0.018\pm0.006$\,d$^{-1}$ is also found. The longer lag feature at a lower frequency would correspond to a more extended emission region while the shorter lag feature at a higher frequency would correspond to a more compact region. This could be explained by the presence of an extended UV emission region on the disc where reverberation happens.

Given that the lags are consistent with zero lag within $2\sigma$ uncertainty ranges, we conclude that only tentative evidence for a broad lag feature is found by applying Gaussian processes to the UVW2 and X-ray lightcurves of \src. Previous attempts to identify lags between two wavelengths of \src\ based on cross-correlation analysis in the time domain suggests similar results \citep[e.g.][]{gallo18}.

\newpage

\section{Conclusions}
\label{conc}

Following the interpolation of the unevenly-sampled lightcurves of \src\ using Gaussian processes, tentative evidence for broad lag features is found in the Fourier frequency domain. The magnitude of the lags is consistent with previous cross-correlation analyses. In addition, the broad lag features {might} suggest an extended emission region e.g. of the disc in \src \: during the reverberation processes. {If the corona is compact within 5 $R_{\rm g}$ in Mrk~335 \citep{2015_Wilkins}, our data suggest a possibly wide range of UVW2 emission radii.}

The structure functions computed from the \textsc{gp}-interpolated lightcurves are consistent with those derived from the observational data and furthermore, illicit potential insights into the properties of the latent lightcurves. In particular, we show through a simulation study that it is possible that dips in the X-ray structure function may be produced by sampling artefacts arising from gaps in the lightcurve. In contrast, the \textsc{gp} structure functions show no dips. While this is not proof that the dip in the observational X-ray structure function is due to a sampling artefact, it does allude to the possibility. The UVW2 \textsc{gp} structure functions do not exhibit strong dependence on the choice of kernel with both Matern $\frac{1}{2}$ and rational quadratic showing up a broken power law with breaks at 139 and 155 days respectively. The X-ray structure functions however do show up differences between kernels with the rational quadratic kernel predicting a power law and the Matern 1/2 kernel predicting a broken power law.

From the Gaussian process modelling perspective, the ability to carry out Bayesian model selection affords a quantitative means of comparing analytic kernels under the marginal likelihood. It may be possible to incorporate further flexibility into the fitting procedure by making use of more sophisticated methods of kernel design \citep{2014_Duvenaud} in order to allow the assessment of fits of sums and products of analytic kernels or by leveraging advances in transforming Gaussian process priors via Deep Gaussian processes \citep{2013_Damianou} or normalising flows \citep{2020_Maronas}. Such approaches could be validated using simulations studies. Additionally, modelling the cross-correlation using multioutput Gaussian processes \citep{2020_de_Wolff} may be an interesting avenue for comparison against the approach taken here. Lastly, Bayesian spectral density estimation \citep{2018_Tobar} may afford further flexibility through nonparametric modelling of the power spectral density in addition to nonparametric modelling of the lightcurve in the time domain. These improvements in Bayesian modelling machinery may help to minimise model misspecification and as such, enable more robust inferences to be made about the functional forms of the latent lightcurves.

\section{Acknowledgements}

Jiachen Jiang acknowledges support from the Tsinghua Shui'Mu Scholar Program and the Tsinghua Astrophysics Outstanding Fellowship.

\clearpage
\bibliography{sample63}

@ARTICLE{pringle81,
       author = {{Pringle}, J.~E.},
        title = "{Accretion discs in astrophysics}",
      journal = {Annual Review of Astronomy and Astrophysics},
         year = 1981,
        month = jan,
       volume = {19},
        pages = {137-162},
          doi = {10.1146/annurev.aa.19.090181.001033},
       adsurl = {https://ui.adsabs.harvard.edu/abs/1981ARA&A..19..137P}
}

@ARTICLE{morgan08,
       author = {{Morgan}, Christopher W. and {Kochanek}, Christopher. S. and
         {Dai}, Xinyu and {Morgan}, Nicholas D. and {Falco}, Emilio E.},
        title = "{X-Ray and Optical Microlensing in the Lensed Quasar PG 1115+080}",
      journal = {The Astrophysical Journal},
         year = 2008,
        month = dec,
       volume = {689},
       number = {2},
        pages = {755-761},
          doi = {10.1086/592767},
archivePrefix = {arXiv},
       eprint = {0802.1210},
 primaryClass = {astro-ph},
       adsurl = {https://ui.adsabs.harvard.edu/abs/2008ApJ...689..755M}
}

@ARTICLE{xin20,
       author = {{Xin}, Chengcheng and {Charisi}, Maria and {Haiman}, Zoltan and {Schiminovich}, David},
        title = "{Correlation between optical and UV variability of a large sample of quasars}",
      journal = {Monthly Notices of the Royal Astronomical Society},
         year = 2020,
        month = jun,
       volume = {495},
       number = {1},
        pages = {1403-1413},
          doi = {10.1093/mnras/staa1258},
archivePrefix = {arXiv},
       eprint = {2001.03154},
 primaryClass = {astro-ph.GA},
       adsurl = {https://ui.adsabs.harvard.edu/abs/2020MNRAS.495.1403X}
}

@ARTICLE{smith18,
       author = {{Smith}, Krista Lynne and {Mushotzky}, Richard F. and {Boyd}, Patricia T. and {Malkan}, Matt and {Howell}, Steve B. and {Gelino}, Dawn M.},
        title = "{The Kepler Light Curves of AGN: A Detailed Analysis}",
      journal = {The Astrophysical Journal},
         year = 2018,
        month = apr,
       volume = {857},
       number = {2},
          eid = {141},
        pages = {141},
          doi = {10.3847/1538-4357/aab88d},
archivePrefix = {arXiv},
       eprint = {1803.06436},
 primaryClass = {astro-ph.HE},
       adsurl = {https://ui.adsabs.harvard.edu/abs/2018ApJ...857..141S}
}

@ARTICLE{sanchez18,
       author = {{S{\'a}nchez-S{\'a}ez}, P. and {Lira}, P. and {Mej{\'\i}a-Restrepo}, J. and {Ho}, L.~C. and {Ar{\'e}valo}, P. and {Kim}, M. and {Cartier}, R. and {Coppi}, P.},
        title = "{The QUEST-La Silla AGN Variability Survey: Connection between AGN Variability and Black Hole Physical Properties}",
      journal = {The Astrophysical Journal},
         year = 2018,
        month = sep,
       volume = {864},
       number = {1},
          eid = {87},
        pages = {87},
          doi = {10.3847/1538-4357/aad7f9},
archivePrefix = {arXiv},
       eprint = {1808.00967},
 primaryClass = {astro-ph.GA},
       adsurl = {https://ui.adsabs.harvard.edu/abs/2018ApJ...864...87S}
}

@ARTICLE{welsh11,
       author = {{Welsh}, B.~Y. and {Wheatley}, J.~M. and {Neil}, J.~D.},
        title = "{GALEX observations of quasar variability in the ultraviolet}",
      journal = {Astronomy and Astrophysics},
         year = 2011,
        month = mar,
       volume = {527},
          eid = {A15},
        pages = {A15},
          doi = {10.1051/0004-6361/201015865},
archivePrefix = {arXiv},
       eprint = {1101.2191},
 primaryClass = {astro-ph.GA},
       adsurl = {https://ui.adsabs.harvard.edu/abs/2011A&A...527A..15W}
}

@ARTICLE{gezari13,
       author = {{Gezari}, S. and {Martin}, D.~C. and {Forster}, K. and {Neill}, J.~D. and {Huber}, M. and {Heckman}, T. and {Bianchi}, L. and {Morrissey}, P. and {Neff}, S.~G. and {Seibert}, M. and {Schiminovich}, D. and {Wyder}, T.~K. and {Burgett}, W.~S. and {Chambers}, K.~C. and {Kaiser}, N. and {Magnier}, E.~A. and {Price}, P.~A. and {Tonry}, J.~L.},
        title = "{The GALEX Time Domain Survey. I. Selection and Classification of Over a Thousand Ultraviolet Variable Sources}",
      journal = {The Astrophysical Journal},
         year = 2013,
        month = mar,
       volume = {766},
       number = {1},
          eid = {60},
        pages = {60},
          doi = {10.1088/0004-637X/766/1/60},
archivePrefix = {arXiv},
       eprint = {1302.1581},
 primaryClass = {astro-ph.CO},
       adsurl = {https://ui.adsabs.harvard.edu/abs/2013ApJ...766...60G}
}

@ARTICLE{jiang21,
       author = {{Jiang}, Jiachen and {Cheng}, Huaqing and {Gallo}, Luigi C. and {Ho}, Luis C. and {Buisson}, Douglas J.~K. and {Fabian}, Andrew C. and {Harrison}, Fiona A. and {Parker}, Michael L. and {Reynolds}, Christopher S. and {Steiner}, James F. and {Tomsick}, John A. and {Walton}, Dominic J. and {Yuan}, Weimin},
        title = "{The awakening beast in the Seyfert 1 Galaxy KUG 1141+371 - I}",
      journal = {Monthly Notices of the Royal Astronomical Society},
         year = 2021,
        month = feb,
       volume = {501},
       number = {1},
        pages = {916-932},
          doi = {10.1093/mnras/staa3737},
archivePrefix = {arXiv},
       eprint = {2011.14254},
 primaryClass = {astro-ph.HE},
       adsurl = {https://ui.adsabs.harvard.edu/abs/2021MNRAS.501..916J}
}

@ARTICLE{cai20,
       author = {{Cai}, Zhen-Yi and {Wang}, Jun-Xian and {Sun}, Mouyuan},
        title = "{EUCLIA. II. On the Puzzling Large UV to X-Ray Lags in Seyfert Galaxies}",
      journal = {The Astrophysical Journal},
         year = 2020,
        month = mar,
       volume = {892},
       number = {1},
          eid = {63},
        pages = {63},
          doi = {10.3847/1538-4357/ab7991},
archivePrefix = {arXiv},
       eprint = {2002.11116},
 primaryClass = {astro-ph.HE},
       adsurl = {https://ui.adsabs.harvard.edu/abs/2020ApJ...892...63C}
}

@ARTICLE{kammoun20,
       author = {{Kammoun}, E.~S. and {Dov{\v{c}}iak}, M. and {Papadakis}, I.~E. and {Caballero-Garc{\'\i}a}, M.~D. and {Karas}, V.},
        title = "{UV/Optical Disk Thermal Reverberation in Active Galactic Nuclei: An In-depth Study with an Analytic Prescription for Time-lag Spectra}",
      journal = {The Astrophysical Journal},
         year = 2021,
        month = jan,
       volume = {907},
       number = {1},
          eid = {20},
        pages = {20},
          doi = {10.3847/1538-4357/abcb93},
archivePrefix = {arXiv},
       eprint = {2011.08563},
 primaryClass = {astro-ph.HE},
       adsurl = {https://ui.adsabs.harvard.edu/abs/2021ApJ...907...20K}
}

@ARTICLE{gardner17,
       author = {{Gardner}, Emma and {Done}, Chris},
        title = "{The origin of the UV/optical lags in NGC 5548}",
      journal = {Monthly Notices of the Royal Astronomical Society},
         year = 2017,
        month = sep,
       volume = {470},
       number = {3},
        pages = {3591-3605},
          doi = {10.1093/mnras/stx946},
archivePrefix = {arXiv},
       eprint = {1603.09564},
 primaryClass = {astro-ph.HE},
       adsurl = {https://ui.adsabs.harvard.edu/abs/2017MNRAS.470.3591G}
}

@ARTICLE{starkey17,
       author = {{Starkey}, D. and {Horne}, Keith and {Fausnaugh}, M.~M. and {Peterson}, B.~M. and {Bentz}, M.~C. and {Kochanek}, C.~S. and {Denney}, K.~D. and {Edelson}, R. and {Goad}, M.~R. and {De Rosa}, G. and {Anderson}, M.~D. and {Ar{\'e}valo}, P. and {Barth}, A.~J. and {Bazhaw}, C. and {Borman}, G.~A. and {Boroson}, T.~A. and {Bottorff}, M.~C. and {Brandt}, W.~N. and {Breeveld}, A.~A. and {Cackett}, E.~M. and {Carini}, M.~T. and {Croxall}, K.~V. and {Crenshaw}, D.~M. and {Dalla Bont{\`a}}, E. and {De Lorenzo-C{\'a}ceres}, A. and {Dietrich}, M. and {Efimova}, N.~V. and {Ely}, J. and {Evans}, P.~A. and {Filippenko}, A.~V. and {Flatland}, K. and {Gehrels}, N. and {Geier}, S. and {Gelbord}, J.~M. and {Gonzalez}, L. and {Gorjian}, V. and {Grier}, C.~J. and {Grupe}, D. and {Hall}, P.~B. and {Hicks}, S. and {Horenstein}, D. and {Hutchison}, T. and {Im}, M. and {Jensen}, J.~J. and {Joner}, M.~D. and {Jones}, J. and {Kaastra}, J. and {Kaspi}, S. and {Kelly}, B.~C. and {Kennea}, J.~A. and {Kim}, S.~C. and {Kim}, M. and {Klimanov}, S.~A. and {Korista}, K.~T. and {Kriss}, G.~A. and {Lee}, J.~C. and {Leonard}, D.~C. and {Lira}, P. and {MacInnis}, F. and {Manne-Nicholas}, E.~R. and {Mathur}, S. and {McHardy}, I.~M. and {Montouri}, C. and {Musso}, R. and {Nazarov}, S.~V. and {Norris}, R.~P. and {Nousek}, J.~A. and {Okhmat}, D.~N. and {Pancoast}, A. and {Parks}, J.~R. and {Pei}, L. and {Pogge}, R.~W. and {Pott}, J. -U. and {Rafter}, S.~E. and {Rix}, H. -W. and {Saylor}, D.~A. and {Schimoia}, J.~S. and {Schn{\"u}lle}, K. and {Sergeev}, S.~G. and {Siegel}, M.~H. and {Spencer}, M. and {Sung}, H. -I. and {Teems}, K.~G. and {Turner}, C.~S. and {Uttley}, P. and {Vestergaard}, M. and {Villforth}, C. and {Weiss}, Y. and {Woo}, J. -H. and {Yan}, H. and {Young}, S. and {Zheng}, W. and {Zu}, Y.},
        title = "{Space Telescope and Optical Reverberation Mapping Project.VI. Reverberating Disk Models for NGC 5548}",
      journal = {The Astrophysical Journal},
         year = 2017,
        month = jan,
       volume = {835},
       number = {1},
          eid = {65},
        pages = {65},
          doi = {10.3847/1538-4357/835/1/65},
archivePrefix = {arXiv},
       eprint = {1611.06051},
 primaryClass = {astro-ph.GA},
       adsurl = {https://ui.adsabs.harvard.edu/abs/2017ApJ...835...65S}
}

@ARTICLE{kara13,
   author = {{Kara}, E. and {Fabian}, A.~C. and {Cackett}, E.~M. and {Uttley}, P. and 
	{Wilkins}, D.~R. and {Zoghbi}, A.},
    title = "{Discovery of high-frequency iron K lags in Ark 564 and Mrk 335}",
  journal = {Monthly Notices of the Royal Astronomical Society},
archivePrefix = "arXiv",
   eprint = {1306.2551},
 primaryClass = "astro-ph.HE",
     year = 2013,
    month = sep,
   volume = 434,
    pages = {1129-1137},
      doi = {10.1093/mnras/stt1055},
   adsurl = {http://adsabs.harvard.edu/abs/2013MNRAS.434.1129K}
}

@ARTICLE{fabian09,
   author = {{Fabian}, A.~C. and {Zoghbi}, A. and {Ross}, R.~R. and {Uttley}, P. and 
	{Gallo}, L.~C. and {Brandt}, W.~N. and {Blustin}, A.~J. and 
	{Boller}, T. and {Caballero-Garcia}, M.~D. and {Larsson}, J. and 
	{Miller}, J.~M. and {Miniutti}, G. and {Ponti}, G. and {Reis}, R.~C. and 
	{Reynolds}, C.~S. and {Tanaka}, Y. and {Young}, A.~J.},
    title = "{Broad line emission from iron K- and L-shell transitions in the active galaxy 1H0707-495}",
  journal = {\nat},
     year = 2009,
    month = may,
   volume = 459,
    pages = {540-542},
      doi = {10.1038/nature08007},
   adsurl = {http://adsabs.harvard.edu/abs/2009Natur.459..540F}
}

@ARTICLE{dexter11,
       author = {{Dexter}, Jason and {Fragile}, P. Chris},
        title = "{Observational Signatures of Tilted Black Hole Accretion Disks from Simulations}",
      journal = {The Astrophysical Journal},
         year = 2011,
        month = mar,
       volume = {730},
       number = {1},
          eid = {36},
        pages = {36},
          doi = {10.1088/0004-637X/730/1/36},
archivePrefix = {arXiv},
       eprint = {1101.3783},
 primaryClass = {astro-ph.HE},
       adsurl = {https://ui.adsabs.harvard.edu/abs/2011ApJ...730...36D}
}

@ARTICLE{zhu16,
       author = {{Zhu}, Fei-Fan and {Wang}, Jun-Xian and {Cai}, Zhen-Yi and {Sun}, Yu-Han},
        title = "{The Timescale-dependent Color Variability of Quasars Viewed with /GALEX}",
      journal = {The Astrophysical Journal},
         year = 2016,
        month = nov,
       volume = {832},
       number = {1},
          eid = {75},
        pages = {75},
          doi = {10.3847/0004-637X/832/1/75},
archivePrefix = {arXiv},
       eprint = {1609.07136},
 primaryClass = {astro-ph.GA},
       adsurl = {https://ui.adsabs.harvard.edu/abs/2016ApJ...832...75Z}
}

@ARTICLE{chartas17,
       author = {{Chartas}, G. and {Krawczynski}, H. and {Zalesky}, L. and
         {Kochanek}, C.~S. and {Dai}, X. and {Morgan}, C.~W. and {Mosquera}, A.},
        title = "{Measuring the Innermost Stable Circular Orbits of Supermassive Black Holes}",
      journal = {The Astrophysical Journal},
         year = 2017,
        month = mar,
       volume = {837},
       number = {1},
          eid = {26},
        pages = {26},
          doi = {10.3847/1538-4357/aa5d50},
archivePrefix = {arXiv},
       eprint = {1609.09490},
 primaryClass = {astro-ph.HE},
       adsurl = {https://ui.adsabs.harvard.edu/abs/2017ApJ...837...26C}
}

@ARTICLE{gaskell03,
       author = {{Gaskell}, C.~M. and {Klimek}, E.~S.},
        title = "{Variability of Active Galactic Nuclei from Optical to X-ray Regions}",
      journal = {Astronomical and Astrophysical Transactions},
         year = 2003,
        month = aug,
       volume = {22},
       number = {4-5},
        pages = {661-680},
          doi = {10.1080/1055679031000153851},
archivePrefix = {arXiv},
       eprint = {0907.1415},
 primaryClass = {astro-ph.CO},
       adsurl = {https://ui.adsabs.harvard.edu/abs/2003A&AT...22..661G}
}

@ARTICLE{shemmer01,
       author = {{Shemmer}, O. and {Romano}, P. and {Bertram}, R. and {Brinkmann}, W. and
         {Collier}, S. and {Crowley}, K.~A. and {Detsis}, E. and
         {Filippenko}, A.~V. and {Gaskell}, C.~M. and {George}, T.~A. and
         {Gliozzi}, M. and {Hiller}, M.~E. and {Jewell}, T.~L. and {Kaspi}, S. and
         {Klimek}, E.~S. and {Lannon}, M.~H. and {Li}, W. and {Martini}, P. and
         {Mathur}, S. and {Negoro}, H. and {Netzer}, H. and {Papadakis}, I. and
         {Papamastorakis}, I. and {Peterson}, B.~M. and {Peterson}, B.~W. and
         {Pogge}, R.~W. and {Pronik}, V.~I. and {Rumstay}, K.~S. and
         {Sergeev}, S.~G. and {Sergeeva}, E.~A. and {Stirpe}, G.~M. and
         {Taylor}, C.~J. and {Treffers}, R.~R. and {Turner}, T.~J. and
         {Uttley}, P. and {Vestergaard}, M. and {von Braun}, K. and
         {Wagner}, R.~M. and {Zheng}, Z.},
        title = "{Multiwavelength Monitoring of the Narrow-Line Seyfert 1 Galaxy Arakelian 564. III. Optical Observations and the Optical-UV-X-Ray Connection}",
      journal = {The Astrophysical Journal},
         year = 2001,
        month = nov,
       volume = {561},
       number = {1},
        pages = {162-170},
          doi = {10.1086/323236},
archivePrefix = {arXiv},
       eprint = {astro-ph/0107075},
 primaryClass = {astro-ph},
       adsurl = {https://ui.adsabs.harvard.edu/abs/2001ApJ...561..162S}
}

@ARTICLE{mushotzky93,
       author = {{Mushotzky}, Richard F. and {Done}, Christine and {Pounds}, Kenneth A.},
        title = "{X-ray spectra and time variability of active galactic nuclei.}",
      journal = {Annual Review of Astronomy and Astrophysics},
         year = 1993,
        month = jan,
       volume = {31},
        pages = {717-717},
          doi = {10.1146/annurev.aa.31.090193.003441},
       adsurl = {https://ui.adsabs.harvard.edu/abs/1993ARA&A..31..717M}
}

@article{2019_Wilkins,
   title={Low-frequency X-ray timing with Gaussian processes and reverberation in the radio-loud AGN 3C 120},
   volume={489},
   ISSN={1365-2966},
   url={http://dx.doi.org/10.1093/mnras/stz2269},
   DOI={10.1093/mnras/stz2269},
   number={2},
   journal={Monthly Notices of the Royal Astronomical Society},
   publisher={Oxford University Press (OUP)},
   author={Wilkins, D R},
   year={2019},
   month={Aug},
   pages={1957–1972}
}

@article{1987_Davies,
  title={Tests for Hurst effect},
  author={Davies, Robert B and Harte, DS},
  journal={Biometrika},
  volume={74},
  number={1},
  pages={95--101},
  year={1987},
  publisher={Oxford University Press}
}

@article{tripathi20,
   title={Tracking the year-to-year variation in the spectral energy distribution of the narrow-line Seyfert 1 galaxy Mrk 335},
   volume={499},
   ISSN={1365-2966},
   url={http://dx.doi.org/10.1093/mnras/staa2817},
   DOI={10.1093/mnras/staa2817},
   number={1},
   journal={Monthly Notices of the Royal Astronomical Society},
   publisher={Oxford University Press (OUP)},
   author={Tripathi, S and McGrath, K M and Gallo, L C and Grupe, D and Komossa, S and Berton, M and Kriss, G and Longinotti, A L},
   year={2020},
   month={Sep},
   pages={1266–1286}
}

@article{1995_Timmer,
  title={On generating power law noise.},
  author={Timmer, J and K{\"o}nig, M},
  journal={Astronomy and Astrophysics},
  volume={300},
  pages={707},
  year={1995}
}

@book{2012_Stein,
  title={Interpolation of spatial data: some theory for kriging},
  author={Stein, Michael L},
  year={2012},
  publisher={Springer Science \& Business Media}
}

@book{2006_Rasmussen,
  author = {Rasmussen, C. E. and Williams, C. K. I.},
  publisher = {MIT Press},
  title = {{G}aussian Processes for Machine Learning},
  year = 2006
}

@article{1992_MacKay_Hierarchical,
  title={Bayesian interpolation},
  author={MacKay, David JC},
  journal={Neural {C}omputation},
  volume={4},
  number={3},
  pages={415--447},
  year={1992},
  publisher={MIT Press}
}

@inproceedings{2001_Rasmussen,
  title={Occam's razor},
  author={Rasmussen, Carl Edward and Ghahramani, Zoubin},
  booktitle={Advances in {N}eural {I}nformation {P}rocessing {S}ystems},
  pages={294--300},
  year={2001}
}

@PHDTHESIS{1992_MacKay,
 AUTHOR         ="D. J. C.  MacKay",
 TITLE          ="Bayesian Methods for Adaptive Models",
 YEAR           =1991,
 SCHOOL		="California Institute of Technology"}

@ARTICLE{GPflow,
   author = {Matthews, Alexander G. de G. and {van der Wilk}, Mark and Nickson, Tom and
	Fujii, Keisuke. and {Boukouvalas}, Alexis and {Le{\'o}n-Villagr{\'a}}, Pablo and
	Ghahramani, Zoubin and Hensman, James},
    title = "{{GP}flow: A {G}aussian process library using {T}ensor{F}low}",
  journal = {Journal of Machine Learning Research},
  year    = {2017},
  month = {apr},
  volume  = {18},
  number  = {40},
  pages   = {1-6},
  url     = {http://jmlr.org/papers/v18/16-537.html}
}

@phdthesis{2014_Duvenaud,
   title = {Automatic Model Construction with {G}aussian Processes},
   author = {David Duvenaud},
   year = {2014},
   school = {{Computational and Biological Learning Laboratory, University of Cambridge}},
}

@phdthesis{2019_der_Wilk,
  title={Sparse {G}aussian process approximations and applications},
  author={van der Wilk, Mark},
  year={2019},
  school={University of Cambridge}
}

@inproceedings{2018_Tobar,
  title={Bayesian nonparametric spectral estimation},
  author={Tobar, Felipe},
  booktitle={Advances in Neural Information Processing Systems},
  pages={10127--10137},
  year={2018}
}

@article{2012_Wang,
  title={Nonparametric Bayesian estimation of periodic light curves},
  author={Wang, Yuyang and Khardon, Roni and Protopapas, Pavlos},
  journal={The Astrophysical Journal},
  volume={756},
  number={1},
  pages={67},
  year={2012},
  publisher={IOP Publishing}
}

@article{2016_Durrande,
  title={Detecting periodicities with Gaussian processes},
  author={Durrande, Nicolas and Hensman, James and Rattray, Magnus and Lawrence, Neil D},
  journal={PeerJ Computer Science},
  volume={2},
  pages={e50},
  year={2016},
  publisher={PeerJ Inc.}
}

@inproceedings{2015_Tobar,
  title={Learning stationary time series using Gaussian processes with nonparametric kernels},
  author={Tobar, Felipe and Bui, Thang D and Turner, Richard E},
  booktitle={Advances in Neural Information Processing Systems},
  pages={3501--3509},
  year={2015}
}

@ARTICLE{shakura73,
       author = {{Shakura}, N.~I. and {Sunyaev}, R.~A.},
        title = "{Reprint of 1973A\&amp;A....24..337S. Black holes in binary systems. Observational appearance.}",
      journal = {Astronomy and Astrophysics},
         year = 1973,
        month = jun,
       volume = {500},
        pages = {33-51},
       adsurl = {https://ui.adsabs.harvard.edu/abs/1973A&A....24..337S}
}

@ARTICLE{mchardy16,
       author = {{McHardy}, I.~M. and {Connolly}, S.~D. and {Peterson}, B.~M. and
         {Bieryla}, A. and {Chand}, H. and {Elvis}, M.~S. and
         {Emmanoulopoulos}, D. and {Falco}, E. and {Gandhi}, P. and {Kaspi}, S. and
         {Latham}, D. and {Lira}, P. and {McCully}, C. and {Netzer}, H. and
         {Uemura}, M.},
        title = "{The origin of UV-optical variability in AGN and test of disc models: XMM-Newton and ground-based observations of NGC 4395}",
      journal = {Astronomische Nachrichten},
         year = 2016,
        month = may,
       volume = {337},
       number = {4-5},
        pages = {500},
          doi = {10.1002/asna.201612337},
archivePrefix = {arXiv},
       eprint = {1601.00215},
 primaryClass = {astro-ph.GA},
       adsurl = {https://ui.adsabs.harvard.edu/abs/2016AN....337..500M}
}

@ARTICLE{smith07,
       author = {{Smith}, R. and {Vaughan}, S.},
        title = "{X-ray and optical variability of Seyfert 1 galaxies as observed with XMM-Newton}",
      journal = {Monthly Notices of the Royal Astronomical Society},
         year = 2007,
        month = mar,
       volume = {375},
       number = {4},
        pages = {1479-1487},
          doi = {10.1111/j.1365-2966.2006.11413.x},
archivePrefix = {arXiv},
       eprint = {astro-ph/0701206},
 primaryClass = {astro-ph},
       adsurl = {https://ui.adsabs.harvard.edu/abs/2007MNRAS.375.1479S}
}

@ARTICLE{grupe07,
       author = {{Grupe}, Dirk and {Komossa}, Stefanie and {Gallo}, Luigi C.},
        title = "{Discovery of the Narrow-Line Seyfert 1 Galaxy Markarian 335 in a Historical Low X-Ray Flux State}",
      journal = {The Astrophysical Journal Letters},
         year = 2007,
        month = oct,
       volume = {668},
       number = {2},
        pages = {L111-L114},
          doi = {10.1086/523042},
archivePrefix = {arXiv},
       eprint = {0709.0733},
 primaryClass = {astro-ph},
       adsurl = {https://ui.adsabs.harvard.edu/abs/2007ApJ...668L.111G}
}

@ARTICLE{grupe12,
       author = {{Grupe}, Dirk and {Komossa}, S. and {Gallo}, Luigi C. and
         {Longinotti}, Anna Lia and {Fabian}, Andrew C. and {Pradhan}, Anil K. and
         {Gruberbauer}, Michael and {Xu}, Dawei},
        title = "{A Remarkable Long-term Light Curve and Deep, Low-state Spectroscopy: Swift and XMM-Newton Monitoring of the NLS1 Galaxy Mkn 335}",
      journal = {The Astrophysical Journal Supplement Series},
         year = 2012,
        month = apr,
       volume = {199},
       number = {2},
          eid = {28},
        pages = {28},
          doi = {10.1088/0067-0049/199/2/28},
archivePrefix = {arXiv},
       eprint = {1202.4692},
 primaryClass = {astro-ph.HE},
       adsurl = {https://ui.adsabs.harvard.edu/abs/2012ApJS..199...28G}
}

@ARTICLE{edelson00,
       author = {{Edelson}, Rick and {Koratkar}, Anuradha and {Nandra}, Kirpal and
         {Goad}, Michael and {Peterson}, Bradley M. and {Collier}, Stefan and
         {Krolik}, Julian and {Malkan}, Matthew and {Maoz}, Dan and
         {O'Brien}, Paul and {Shull}, J. Michael and {Vaughan}, Simon and
         {Warwick}, Robert},
        title = "{Intensive HST, RXTE, and ASCA Monitoring of NGC 3516: Evidence against Thermal Reprocessing}",
      journal = {The Astrophysical Journal},
         year = 2000,
        month = may,
       volume = {534},
       number = {1},
        pages = {180-188},
          doi = {10.1086/308752},
archivePrefix = {arXiv},
       eprint = {astro-ph/9912266},
 primaryClass = {astro-ph},
       adsurl = {https://ui.adsabs.harvard.edu/abs/2000ApJ...534..180E}
}

@ARTICLE{troyer16,
       author = {{Troyer}, Jon and {Starkey}, David and {Cackett}, Edward M. and
         {Bentz}, Misty C. and {Goad}, Michael R. and {Horne}, Keith and
         {Seals}, James E.},
        title = "{Correlated X-ray/ultraviolet/optical variability in NGC 6814}",
      journal = {Monthly Notices of the Royal Astronomical Society},
         year = 2016,
        month = mar,
       volume = {456},
       number = {4},
        pages = {4040-4050},
          doi = {10.1093/mnras/stv2862},
archivePrefix = {arXiv},
       eprint = {1509.01124},
 primaryClass = {astro-ph.GA},
       adsurl = {https://ui.adsabs.harvard.edu/abs/2016MNRAS.456.4040T}
}

@ARTICLE{shappee14,
       author = {{Shappee}, B.~J. and {Prieto}, J.~L. and {Grupe}, D. and
         {Kochanek}, C.~S. and {Stanek}, K.~Z. and {De Rosa}, G. and
         {Mathur}, S. and {Zu}, Y. and {Peterson}, B.~M. and {Pogge}, R.~W. and
         {Komossa}, S. and {Im}, M. and {Jencson}, J. and {Holoien}, T.~W. -S. and
         {Basu}, U. and {Beacom}, J.~F. and {Szczygie{\l}}, D.~M. and
         {Brimacombe}, J. and {Adams}, S. and {Campillay}, A. and {Choi}, C. and
         {Contreras}, C. and {Dietrich}, M. and {Dubberley}, M. and
         {Elphick}, M. and {Foale}, S. and {Giustini}, M. and {Gonzalez}, C. and
         {Hawkins}, E. and {Howell}, D.~A. and {Hsiao}, E.~Y. and {Koss}, M. and
         {Leighly}, K.~M. and {Morrell}, N. and {Mudd}, D. and {Mullins}, D. and
         {Nugent}, J.~M. and {Parrent}, J. and {Phillips}, M.~M. and
         {Pojmanski}, G. and {Rosing}, W. and {Ross}, R. and {Sand}, D. and
         {Terndrup}, D.~M. and {Valenti}, S. and {Walker}, Z. and {Yoon}, Y.},
        title = "{The Man behind the Curtain: X-Rays Drive the UV through NIR Variability in the 2013 Active Galactic Nucleus Outburst in NGC 2617}",
      journal = {The Astrophysical Journal},
         year = 2014,
        month = jun,
       volume = {788},
       number = {1},
          eid = {48},
        pages = {48},
          doi = {10.1088/0004-637X/788/1/48},
archivePrefix = {arXiv},
       eprint = {1310.2241},
 primaryClass = {astro-ph.HE},
       adsurl = {https://ui.adsabs.harvard.edu/abs/2014ApJ...788...48S}
}

@ARTICLE{gallo18,
       author = {{Gallo}, L.~C. and {Blue}, D.~M. and {Grupe}, D. and {Komossa}, S. and
         {Wilkins}, D.~R.},
        title = "{Eleven years of monitoring the Seyfert 1 Mrk 335 with Swift: Characterizing the X-ray and UV/optical variability}",
      journal = {Monthly Notices of the Royal Astronomical Society},
         year = 2018,
        month = aug,
       volume = {478},
       number = {2},
        pages = {2557-2568},
          doi = {10.1093/mnras/sty1134},
archivePrefix = {arXiv},
       eprint = {1805.00300},
 primaryClass = {astro-ph.HE},
       adsurl = {https://ui.adsabs.harvard.edu/abs/2018MNRAS.478.2557G}
}

@ARTICLE{buisson17,
       author = {{Buisson}, D.~J.~K. and {Lohfink}, A.~M. and {Alston}, W.~N. and
         {Fabian}, A.~C.},
        title = "{Ultraviolet and X-ray variability of active galactic nuclei with Swift}",
      journal = {Monthly Notices of the Royal Astronomical Society},
         year = 2017,
        month = jan,
       volume = {464},
       number = {3},
        pages = {3194-3218},
          doi = {10.1093/mnras/stw2486},
archivePrefix = {arXiv},
       eprint = {1609.08638},
 primaryClass = {astro-ph.HE},
       adsurl = {https://ui.adsabs.harvard.edu/abs/2017MNRAS.464.3194B}
}

@article{2015_Rajpaul,
  title={A {G}aussian process framework for modelling stellar activity signals in radial velocity data},
  author={Rajpaul, Vinesh and Aigrain, Suzanne and Osborne, Michael A and Reece, Steven and Roberts, S},
  journal={Monthly Notices of the Royal Astronomical Society},
  volume={452},
  number={3},
  pages={2269--2291},
  year={2015},
  publisher={Oxford University Press}
}

@phdthesis{2015_Karamanavis,
  title={Zooming into $\gamma$-ray loud galactic nuclei: broadband emission and structure dynamics of the blazar PKS 1502+ 106 and the narrow-line Seyfert 1 1H 0323+ 342},
  author={Karamanavis, Vasileios Vassilis},
  year={2015},
  school={Universit{\"a}t zu K{\"o}ln}
}

@article{2017_Karamanavis,
  title={{G}aussian processes for blazar variability studies},
  author={Karamanavis, Vassilis},
  journal={Galaxies},
  volume={5},
  number={1},
  pages={19},
  year={2017},
  publisher={Multidisciplinary Digital Publishing Institute}
}

@article{2013_Roberts,
  title={Gaussian processes for time-series modelling},
  author={Roberts, Stephen and Osborne, Michael and Ebden, Mark and Reece, Steven and Gibson, Neale and Aigrain, Suzanne},
  journal={Philosophical Transactions of the Royal Society A: Mathematical, Physical and Engineering Sciences},
  volume={371},
  number={1984},
  pages={20110550},
  year={2013},
  publisher={The Royal Society Publishing}
}

@article{2005_Uttley,
  title={X-ray variability of NGC 3227 and 5506 and the nature of active galactic nucleus ‘states’},
  author={Uttley, Philip and McHardy, Ian M},
  journal={Monthly Notices of the Royal Astronomical Society},
  volume={363},
  number={2},
  pages={586--596},
  year={2005},
  publisher={Blackwell Science Ltd Oxford, UK}
}

@article{2019_Parker,
   title={The nuclear environment of the NLS1 Mrk 335: Obscuration of the X-ray line emission by a variable outflow},
   volume={490},
   ISSN={1365-2966},
   url={http://dx.doi.org/10.1093/mnras/stz2566},
   DOI={10.1093/mnras/stz2566},
   number={1},
   journal={Monthly Notices of the Royal Astronomical Society},
   publisher={Oxford University Press (OUP)},
   author={Parker, M L and Longinotti, A L and Schartel, N and Grupe, D and Komossa, S and Kriss, G and Fabian, A C and Gallo, L and Harrison, F A and Jiang, J and et al.},
   year={2019},
   month={Sep},
   pages={683–697}
}

@article{2013_Zoghbi,
  title={Calculating time lags from unevenly sampled light curves},
  author={Zoghbi, Abderahmen and Reynolds, Chris and Cackett, EM},
  journal={The Astrophysical Journal},
  volume={777},
  number={1},
  pages={24},
  year={2013},
  publisher={IOP Publishing}
}

@article{1998_Bond,
  title={Estimating the power spectrum of the cosmic microwave background},
  author={Bond, JR and Jaffe, Andrew H and Knox, L},
  journal={Physical Review D},
  volume={57},
  number={4},
  pages={2117},
  year={1998},
  publisher={APS}
}

@article{2000_Reynolds,
  title={On the Lack of X-Ray Iron Line Reverberation in MCG--6-30-15: Implications for the Black Hole Mass and Accretion Disk Structure},
  author={Reynolds, Christopher S},
  journal={The Astrophysical Journal},
  volume={533},
  number={2},
  pages={811},
  year={2000},
  publisher={IOP Publishing}
}

@article{2010_Miller,
  title={Spectral variability and reverberation time delays in the Suzaku X-ray spectrum of NGC 4051},
  author={Miller, L and Turner, TJ and Reeves, JN and Lobban, A and Kraemer, SB and Crenshaw, DM},
  journal={Monthly Notices of the Royal Astronomical Society},
  volume={403},
  number={1},
  pages={196--210},
  year={2010},
  publisher={Blackwell Publishing Ltd Oxford, UK}
}

@article{1992_Press,
  title={The time delay of gravitational lens 0957+ 561. I-Methodology and analysis of optical photometric data. II-Analysis of radio data and combined optical-radio analysis},
  author={Press, William H and Rybicki, George B and Hewitt, Jacqueline N},
  journal={The Astrophysical Journal},
  volume={385},
  pages={404--420},
  year={1992}
}

@article{1992_Hughes,
  title={The University of Michigan radio astronomy data base. I-Structure function analysis and the relation between BL Lacertae objects and quasi-stellar objects},
  author={Hughes, PA and Aller, HD and Aller, MF},
  journal={The Astrophysical Journal},
  volume={396},
  pages={469--486},
  year={1992}
}

@article{2001_Collier,
  title={Characteristic ultraviolet/optical timescales in active galactic nuclei},
  author={Collier, Stefan and Peterson, Bradley M},
  journal={The Astrophysical Journal},
  volume={555},
  number={2},
  pages={775},
  year={2001},
  publisher={IOP Publishing}
}

@article{1996_di_Clemente,
  title={The Variability of Quasars. II. Frequency Dependence},
  author={di Clemente, A and Giallongo, E and Natali, G and Trevese, D and Vagnetti, F},
  journal={The Astrophysical Journal},
  volume={463},
  pages={466},
  year={1996}
}

@article{1985_Simonetti,
  title={Flicker of extragalactic radio sources at two frequencies},
  author={Simonetti, JH and Cordes, JM and Heeschen, DS},
  journal={The Astrophysical Journal},
  volume={296},
  pages={46--59},
  year={1985}
}

@article{2020_de_Wolff,
title = {MOGPTK: The multi-output {G}aussian process toolkit},
journal = {Neurocomputing},
volume = {424},
pages = {49-53},
year = {2021},
issn = {0925-2312},
doi = {https://doi.org/10.1016/j.neucom.2020.09.085},
author = {Taco {de Wolff} and Alejandro Cuevas and Felipe Tobar}
}

@article{2020_Lyon,
  title={Imbalance Learning for Variable Star Classification},
  author={Lyon, Robert and Hosenie, Zafiirah and Mootoovaloo, Arrykrishna and Stappers, BW and McBride, Vanessa},
  journal={Monthly Notices of the Royal Astronomical Society},
  volume={493},
  number={4},
  pages={6050--6059},
  year={2020},
  publisher={Oxford University Press}
}

@INPROCEEDINGS{2018_Gallo_New,
       author = {{Gallo}, L.},
        title = "{X-ray perspective of Narrow-line Seyfert 1 galaxies}",
    booktitle = {Revisiting Narrow-Line Seyfert 1 Galaxies and their Place in the Universe},
         year = 2018,
        month = apr,
          eid = {34},
        pages = {34},
archivePrefix = {arXiv},
       eprint = {1807.09838},
 primaryClass = {astro-ph.HE},
       adsurl = {https://ui.adsabs.harvard.edu/abs/2018rnls.confE..34G}
}

@ARTICLE{2013_Gallo_New,
       author = {{Gallo}, L.~C. and {Fabian}, A.~C. and {Grupe}, D. and {Bonson}, K. and
         {Komossa}, S. and {Longinotti}, A.~L. and {Miniutti}, G. and
         {Walton}, D.~J. and {Zoghbi}, A. and {Mathur}, S.},
        title = "{A blurred reflection interpretation for the intermediate flux state in Mrk 335}",
      journal = {Monthly Notices of the Royal Astronomical Society},
         year = 2013,
        month = jan,
       volume = {428},
       number = {2},
        pages = {1191-1200},
          doi = {10.1093/mnras/sts102},
archivePrefix = {arXiv},
       eprint = {1210.0855},
 primaryClass = {astro-ph.HE},
       adsurl = {https://ui.adsabs.harvard.edu/abs/2013MNRAS.428.1191G}
}

@ARTICLE{2015_Gallo_New,
       author = {{Gallo}, L.~C. and {Wilkins}, D.~R. and {Bonson}, K. and
         {Chiang}, C. -Y. and {Grupe}, D. and {Parker}, M.~L. and {Zoghbi}, A. and
         {Fabian}, A.~C. and {Komossa}, S. and {Longinotti}, A.~L.},
        title = "{Suzaku observations of Mrk 335: confronting partial covering and relativistic reflection}",
      journal = {Monthly Notices of the Royal Astronomical Society},
         year = 2015,
        month = jan,
       volume = {446},
       number = {1},
        pages = {633-650},
          doi = {10.1093/mnras/stu2108},
archivePrefix = {arXiv},
       eprint = {1410.2330},
 primaryClass = {astro-ph.HE},
       adsurl = {https://ui.adsabs.harvard.edu/abs/2015MNRAS.446..633G}
}

@ARTICLE{2014_Parker_New,
       author = {{Parker}, M.~L. and {Wilkins}, D.~R. and {Fabian}, A.~C. and
         {Grupe}, D. and {Dauser}, T. and {Matt}, G. and {Harrison}, F.~A. and
         {Brenneman}, L. and {Boggs}, S.~E. and {Christensen}, F.~E. and
         {Craig}, W.~W. and {Gallo}, L.~C. and {Hailey}, C.~J. and {Kara}, E. and
         {Komossa}, S. and {Marinucci}, A. and {Miller}, J.~M. and
         {Risaliti}, G. and {Stern}, D. and {Walton}, D.~J. and {Zhang}, W.~W.},
        title = "{The NuSTAR spectrum of Mrk 335: extreme relativistic effects within two gravitational radii of the event horizon?}",
      journal = {Monthly Notices of the Royal Astronomical Society},
         year = 2014,
        month = sep,
       volume = {443},
       number = {2},
        pages = {1723-1732},
          doi = {10.1093/mnras/stu1246},
archivePrefix = {arXiv},
       eprint = {1407.8223},
 primaryClass = {astro-ph.HE},
       adsurl = {https://ui.adsabs.harvard.edu/abs/2014MNRAS.443.1723P}
}

@ARTICLE{2013_Longinotti,
       author = {{Longinotti}, A.~L. and {Krongold}, Y. and {Kriss}, G.~A. and {Ely}, J. and
         {Gallo}, L. and {Grupe}, D. and {Komossa}, S. and {Mathur}, S. and
         {Pradhan}, A.},
        title = "{The Rise of an Ionized Wind in the Narrow-line Seyfert 1 Galaxy Mrk 335 Observed by XMM-Newton and HST}",
      journal = {The Astrophysical Journal},
         year = 2013,
        month = apr,
       volume = {766},
       number = {2},
          eid = {104},
        pages = {104},
          doi = {10.1088/0004-637X/766/2/104},
archivePrefix = {arXiv},
       eprint = {1301.5463},
 primaryClass = {astro-ph.CO},
       adsurl = {https://ui.adsabs.harvard.edu/abs/2013ApJ...766..104L}
}

@ARTICLE{2019_Longinotti,
       author = {{Longinotti}, Anna Lia and {Kriss}, Gerard and {Krongold}, Yair and
         {Arellano-Cordova}, Karla Z. and {Komossa}, S. and {Gallo}, Luigi and
         {Grupe}, Dirk and {Mathur}, Smita and {Parker}, Michael L. and
         {Pradhan}, Anil and {Wilkins}, Dan},
        title = "{The XMM-Newton/HST View of the Obscuring Outflow in the Seyfert Galaxy Mrk 335 Observed at Extremely Low X-Ray Flux}",
      journal = {The Astrophysical Journal},
         year = 2019,
        month = apr,
       volume = {875},
       number = {2},
          eid = {150},
        pages = {150},
          doi = {10.3847/1538-4357/ab125a},
archivePrefix = {arXiv},
       eprint = {1903.05795},
 primaryClass = {astro-ph.GA},
       adsurl = {https://ui.adsabs.harvard.edu/abs/2019ApJ...875..150L}
}

@ARTICLE{2015_Wilkins,
       author = {{Wilkins}, D.~R. and {Gallo}, L.~C. and {Grupe}, D. and {Bonson}, K. and
         {Komossa}, S. and {Fabian}, A.~C.},
        title = "{Flaring from the supermassive black hole in Mrk 335 studied with Swift and NuSTAR}",
      journal = {Monthly Notices of the Royal Astronomical Society},
         year = 2015,
        month = dec,
       volume = {454},
       number = {4},
        pages = {4440-4451},
          doi = {10.1093/mnras/stv2130},
archivePrefix = {arXiv},
       eprint = {1510.07656},
 primaryClass = {astro-ph.HE},
       adsurl = {https://ui.adsabs.harvard.edu/abs/2015MNRAS.454.4440W}
}

@article{2017_Jones,
  title={Improving Exoplanet Detection Power: Multivariate Gaussian Process Models for Stellar Activity},
  author={Jones, David E and Stenning, David C and Ford, Eric B and Wolpert, Robert L and Loredo, Thomas J and Gilbertson, Christian and Dumusque, Xavier},
  journal={arXiv preprint arXiv:1711.01318},
  year={2017}
}

@article{2017_Czekala,
  title={Disentangling Time-series Spectra with Gaussian Processes: Applications to Radial Velocity Analysis},
  author={Czekala, Ian and Mandel, Kaisey S and Andrews, Sean M and Dittmann, Jason A and Ghosh, Sujit K and Montet, Benjamin T and Newton, Elisabeth R},
  journal={The Astrophysical Journal},
  volume={840},
  number={1},
  pages={49},
  year={2017},
  publisher={IOP Publishing}
}

@article{2020_Gordon,
  title={A Fast, Two-dimensional Gaussian Process Method Based on Celerite: Applications to Transiting Exoplanet Discovery and Characterization},
  author={Gordon, Tyler A and Agol, Eric and Foreman-Mackey, Daniel},
  journal={The Astronomical Journal},
  volume={160},
  number={5},
  pages={240},
  year={2020},
  publisher={IOP Publishing}
}

@article{2018_Angus,
  title={Inferring probabilistic stellar rotation periods using Gaussian processes},
  author={Angus, Ruth and Morton, Timothy and Aigrain, Suzanne and Foreman-Mackey, Daniel and Rajpaul, Vinesh},
  journal={Monthly Notices of the Royal Astronomical Society},
  volume={474},
  number={2},
  pages={2094--2108},
  year={2018},
  publisher={Oxford University Press}
}

@article{2020_Yang,
   title={Gaussian Process Modeling Fermi-LAT $\gamma$-Ray Blazar Variability: A Sample of Blazars with $\gamma$-Ray Quasi-periodicities},
   volume={907},
   ISSN={1538-4357},
   url={http://dx.doi.org/10.3847/1538-4357/abcbff},
   DOI={10.3847/1538-4357/abcbff},
   number={2},
   journal={The Astrophysical Journal},
   publisher={American Astronomical Society},
   author={Yang, Shenbang and Yan, Dahai and Zhang, Pengfei and Dai, Benzhong and Zhang, Li},
   year={2021},
   month={Feb},
   pages={105}
}

@article{2020_Covino,
  title={Looking at Blazar Light-curve Periodicities with Gaussian Processes},
  author={Covino, Stefano and Landoni, Marco and Sandrinelli, Angela and Treves, Aldo},
  journal={The Astrophysical Journal},
  volume={895},
  number={2},
  pages={122},
  year={2020},
  publisher={IOP Publishing}
}

@article{2019_Pass,
  title={Estimating dayside effective temperatures of hot Jupiters and associated uncertainties through Gaussian process regression},
  author={Pass, Emily K and Cowan, Nicolas B and Cubillos, Patricio E and Sklar, Jack G},
  journal={Monthly Notices of the Royal Astronomical Society},
  volume={489},
  number={1},
  pages={941--950},
  year={2019},
  publisher={Oxford University Press}
}

@article{2020_Diamond,
  title={Simultaneous Optical Transmission Spectroscopy of a Terrestrial, Habitable-zone Exoplanet with Two Ground-based Multiobject Spectrographs},
  author={Diamond-Lowe, Hannah and Berta-Thompson, Zachory and Charbonneau, David and Dittmann, Jason and Kempton, Eliza M-R},
  journal={The Astronomical Journal},
  volume={160},
  number={1},
  pages={27},
  year={2020},
  publisher={IOP Publishing}
}

@article{2020_Langellier,
   title={Detection Limits of Low-mass, Long-period Exoplanets Using Gaussian Processes Applied to HARPS-N Solar Radial Velocities},
   volume={161},
   ISSN={1538-3881},
   url={http://dx.doi.org/10.3847/1538-3881/abf1e0},
   DOI={10.3847/1538-3881/abf1e0},
   number={6},
   journal={The Astronomical Journal},
   publisher={American Astronomical Society},
   author={Langellier, N. and Milbourne, T. W. and Phillips, D. F. and Haywood, R. D. and Saar, S. H. and Mortier, A. and Malavolta, L. and Thompson, S. and Cameron, A. Collier and Dumusque, X. and et al.},
   year={2021},
   month={May},
   pages={287}
}

@article{2012_Gibson,
  title={A Gaussian process framework for modelling instrumental systematics: application to transmission spectroscopy},
  author={Gibson, NP and Aigrain, Suzanne and Roberts, S and Evans, TM and Osborne, Michael and Pont, F},
  journal={Monthly Notices of the Royal Astronomical Society},
  volume={419},
  number={3},
  pages={2683--2694},
  year={2012},
  publisher={The Royal Astronomical Society}
}

@article{2018_Nikolov,
  title={An absolute sodium abundance for a cloud-free ‘hot Saturn’exoplanet},
  author={Nikolov, Nikolay and Sing, David K and Fortney, Jonathan J and Goyal, Jayesh M and Drummond, Benjamin and Evans, Tom M and Gibson, Neale P and De Mooij, Ernst JW and Rustamkulov, Zafar and Wakeford, Hannah R and others},
  journal={Nature},
  volume={557},
  number={7706},
  pages={526--529},
  year={2018},
  publisher={Nature Publishing Group}
}

@article{2016_Aigrain,
  title={K2SC: flexible systematics correction and detrending of K2 light curves using Gaussian process regression},
  author={Aigrain, Suzanne and Parviainen, Hannu and Pope, BJS},
  journal={Monthly Notices of the Royal Astronomical Society},
  volume={459},
  number={3},
  pages={2408--2419},
  year={2016},
  publisher={Oxford University Press}
}

@misc{2008_Murray,
  title={Introduction to {G}aussian processes},
  author={Murray, Iain},
  year          = {2008},
  publisher={University of Toronto}
}

@article{2018_Buisson,
  title={Is there a UV/X-ray connection in IRAS 13224- 3809?},
  author={Buisson, DJK and Lohfink, AM and Alston, WN and Cackett, EM and Chiang, CY and Dauser, T and De Marco, B and Fabian, AC and Gallo, LC and Garcia, JA and others},
  journal={Monthly Notices of the Royal Astronomical Society},
  volume={475},
  number={2},
  pages={2306--2313},
  year={2018},
  publisher={Oxford University Press}
}

@inproceedings{2020_Maronas,
  title={Transforming {G}aussian processes with normalizing flows},
  author={Maro{\~n}as, Juan and Hamelijnck, Oliver and Knoblauch, Jeremias and Damoulas, Theodoros},
  booktitle={International Conference on Artificial Intelligence and Statistics},
  pages={1081--1089},
  year={2021},
  organization={PMLR}
}

@inproceedings{2013_Damianou,
  title={Deep {G}aussian processes},
  author={Damianou, Andreas and Lawrence, Neil D},
  booktitle={Artificial Intelligence and Statistics},
  pages={207--215},
  year={2013},
  organization={PMLR}
}

@article{2020_Komossa,
  title={Lifting the curtain: The Seyfert galaxy Mrk 335 emerges from deep low-state in a sequence of rapid flare events},
  author={Komossa, S and Grupe, D and Gallo, LC and Poulos, P and Blue, D and Kara, E and Kriss, G and Longinotti, AL and Parker, ML and Wilkins, D},
  journal={Astronomy \& Astrophysics},
  volume={643},
  pages={L7},
  year={2020},
  publisher={EDP Sciences}
}

@article{2020_Greeff,
  title={Exploiting differential flatness for robust learning-based tracking control using {G}aussian processes},
  author={Greeff, Melissa and Schoellig, Angela P},
  journal={IEEE Control Systems Letters},
  volume={5},
  number={4},
  pages={1121--1126},
  year={2020},
  publisher={IEEE}
}

@inproceedings{2011_Deisenroth,
  title={PILCO: {A} model-based and data-efficient approach to policy search},
  author={Deisenroth, Marc and Rasmussen, Carl E},
  booktitle={Proceedings of the 28th International Conference on machine learning (ICML-11)},
  pages={465--472},
  year={2011},
  organization={Citeseer}
}

@ARTICLE{2021_Nigam,
       author = {{Nigam}, AkshatKumar and {Pollice}, Robert and {Hurley}, Matthew F.~D. and {Hickman}, Riley J. and {Aldeghi}, Matteo and {Yoshikawa}, Naruki and {Chithrananda}, Seyone and {Voelz}, Vincent A. and {Aspuru-Guzik}, Al{\'a}n},
        title = "{Assigning Confidence to Molecular Property Prediction}",
      journal = {arXiv e-prints},
         year = 2021,
        month = feb,
          eid = {arXiv:2102.11439},
        pages = {arXiv:2102.11439},
archivePrefix = {arXiv},
       eprint = {2102.11439},
 primaryClass = {cs.LG},
       adsurl = {https://ui.adsabs.harvard.edu/abs/2021arXiv210211439N}
}

@article{2020_Moss,
  title={BOSS: {B}ayesian Optimization over String Spaces},
  author={Moss, Henry and Leslie, David and Beck, Daniel and Gonzalez, Javier and Rayson, Paul},
  journal={Advances in Neural Information Processing Systems},
  volume={33},
  year={2020}
}

@ARTICLE{2020_flowmo,
       author = {{Moss}, Henry B. and {Griffiths}, Ryan-Rhys},
        title = "{Gaussian Process Molecule Property Prediction with FlowMO}",
      journal = {arXiv e-prints},
         year = 2020,
        month = oct,
          eid = {arXiv:2010.01118},
        pages = {arXiv:2010.01118},
archivePrefix = {arXiv},
       eprint = {2010.01118},
 primaryClass = {cs.LG},
       adsurl = {https://ui.adsabs.harvard.edu/abs/2020arXiv201001118M}
}

@ARTICLE{2020_Thawani,
       author = {{Thawani}, Aditya R. and {Griffiths}, Ryan-Rhys and {Jamasb}, Arian and {Bourached}, Anthony and {Jones}, Penelope and {McCorkindale}, William and {Aldrick}, Alexander A. and {Lee}, Alpha A.},
        title = "{The Photoswitch Dataset: A Molecular Machine Learning Benchmark for the Advancement of Synthetic Chemistry}",
      journal = {arXiv e-prints},
         year = 2020,
        month = jun,
          eid = {arXiv:2008.03226},
        pages = {arXiv:2008.03226},
archivePrefix = {arXiv},
       eprint = {2008.03226},
 primaryClass = {physics.chem-ph},
       adsurl = {https://ui.adsabs.harvard.edu/abs/2020arXiv200803226T}
}

@article{2020_Griffiths,
  title={Constrained {B}ayesian optimization for automatic chemical design using variational autoencoders},
  author={Griffiths, Ryan-Rhys and Hern{\'a}ndez-Lobato, Jos{\'e} Miguel},
  journal={Chemical Science},
  volume={11},
  number={2},
  pages={577--586},
  year={2020},
  publisher={Royal Society of Chemistry}
}

@ARTICLE{2020_Grosnit,
       author = {{Grosnit}, Antoine and {Cowen-Rivers}, Alexander I. and {Tutunov}, Rasul and {Griffiths}, Ryan-Rhys and {Wang}, Jun and {Bou-Ammar}, Haitham},
        title = "{Are we Forgetting about Compositional Optimisers in Bayesian Optimisation?}",
      journal = {arXiv e-prints},
         year = 2020,
        month = dec,
          eid = {arXiv:2012.08240},
        pages = {arXiv:2012.08240},
archivePrefix = {arXiv},
       eprint = {2012.08240},
 primaryClass = {cs.LG},
       adsurl = {https://ui.adsabs.harvard.edu/abs/2020arXiv201208240G}
}

@article{2015_Shahriari,
  title={Taking the human out of the loop: {A} review of {B}ayesian optimization},
  author={Shahriari, Bobak and Swersky, Kevin and Wang, Ziyu and Adams, Ryan P and De Freitas, Nando},
  journal={Proceedings of the IEEE},
  volume={104},
  number={1},
  pages={148--175},
  year={2015},
  publisher={IEEE}
}

@ARTICLE{2021_Rivers,
       author = {{Cowen-Rivers}, Alexander I. and {Lyu}, Wenlong and {Tutunov}, Rasul and {Wang}, Zhi and {Grosnit}, Antoine and { Griffiths}, Ryan-Rhys and {Jianye}, Hao and {Wang}, Jun and {Peters}, Jan and {Bou-Ammar}, Haitham},
        title = "{An Empirical Study of Assumptions in Bayesian Optimisation}",
      journal = {arXiv e-prints},
         year = 2021,
          eid = {arXiv:2012.03826},
        pages = {arXiv:2012.03826},
archivePrefix = {arXiv},
       eprint = {2012.03826},
 primaryClass = {cs.LG}
}

@ARTICLE{2019_Griffiths,
       author = {{Griffiths}, Ryan-Rhys and {Aldrick}, Alexander A. and {Garcia-Ortegon}, Miguel and {Lalchand}, Vidhi R. and {Lee}, Alpha A.},
        title = "{Achieving Robustness to Aleatoric Uncertainty with Heteroscedastic Bayesian Optimisation}",
      journal = {arXiv e-prints},
         year = 2019,
        month = oct,
          eid = {arXiv:1910.07779},
        pages = {arXiv:1910.07779},
archivePrefix = {arXiv},
       eprint = {1910.07779},
 primaryClass = {stat.ML},
       adsurl = {https://ui.adsabs.harvard.edu/abs/2019arXiv191007779G}
}

@article{2004_Mchardy,
  title={Combined long and short time-scale X-ray variability of NGC 4051 with RXTE and XMM-Newton},
  author={I. Mchardy and I. Papadakis and P. Uttley and M. J. Page and K. Mason},
  journal={Monthly Notices of the Royal Astronomical Society},
  year={2004},
  volume={348},
  pages={783-801}
}

@ARTICLE{2020_Hase,
       author = {{Hase}, Florian and {Aldeghi}, Matteo and {Hickman}, Riley J. and {Roch}, Loic M. and {Aspuru-Guzik}, Alan},
        title = "{Gryffin: An algorithm for Bayesian optimization of categorical variables informed by expert knowledge}",
      journal = {arXiv e-prints},
         year = 2020,
        month = mar,
          eid = {arXiv:2003.12127},
        pages = {arXiv:2003.12127},
archivePrefix = {arXiv},
       eprint = {2003.12127},
 primaryClass = {stat.ML},
       adsurl = {https://ui.adsabs.harvard.edu/abs/2020arXiv200312127H}
}

@article{2021_Luger1,
  author        = {{Luger}, Rodrigo and {Foreman-Mackey}, Daniel and {Hedges}, Christina and {Hogg}, David W.},
  title         = {{Mapping stellar surfaces I: Degeneracies in the rotational light curve problem}},
  journal       = {arXiv e-prints},
  year          = 2021,
  month         = jan,
  eid           = {arXiv:2102.00007},
  pages         = {arXiv:2102.00007},
  archiveprefix = {arXiv},
  eprint        = {2102.00007},
  primaryclass  = {astro-ph.SR},
  adsurl        = {https://ui.adsabs.harvard.edu/abs/2021arXiv210200007L}
}

@article{2021_Luger2,
  author        = {{Luger}, Rodrigo and {Foreman-Mackey}, Daniel and {Hedges}, Christina},
  title         = {{Mapping stellar surfaces II: An interpretable Gaussian process model for light curves}},
  journal       = {arXiv e-prints},
  year          = 2021,
  month         = feb,
  eid           = {arXiv:2102.01697},
  pages         = {arXiv:2102.01697},
  archiveprefix = {arXiv},
  eprint        = {2102.01697},
  primaryclass  = {astro-ph.SR},
  adsurl        = {https://ui.adsabs.harvard.edu/abs/2021arXiv210201697L}
}

@article{2021_Luger3,
  author        = {{Luger}, Rodrigo and {Foreman-Mackey}, Daniel and {Hedges}, Christina},
  title         = {{starry\_process: Interpretable Gaussian processes for stellar light curves}},
  journal       = {arXiv e-prints},
  year          = 2021,
  month         = feb,
  eid           = {arXiv:2102.01774},
  pages         = {arXiv:2102.01774},
  archiveprefix = {arXiv},
  eprint        = {2102.01774},
  primaryclass  = {astro-ph.SR},
  adsurl        = {https://ui.adsabs.harvard.edu/abs/2021arXiv210201774L}
}

@INPROCEEDINGS{2021_Yu,
       author = {{Yu}, W. and {Richards}, G.~T.},
        title = "{Accelerating CARMA modeling with Gaussian Processes}",
    booktitle = {American Astronomical Society Meeting Abstracts},
         year = 2021,
       series = {American Astronomical Society Meeting Abstracts},
       volume = {53},
        month = jan,
          eid = {541.08},
        pages = {541.08},
       adsurl = {https://ui.adsabs.harvard.edu/abs/2021AAS...23754108Y}
}
\bibliographystyle{agsm}

\appendix

\section{Additional Graphical Tests for Identifying the Flux Distribution}
\label{dist_tests}

\begin{figure}[h!]
\centering
\subfigure[PP plot for X-ray log count rates]{\label{fig:4pt1}\includegraphics[width=0.49\textwidth]{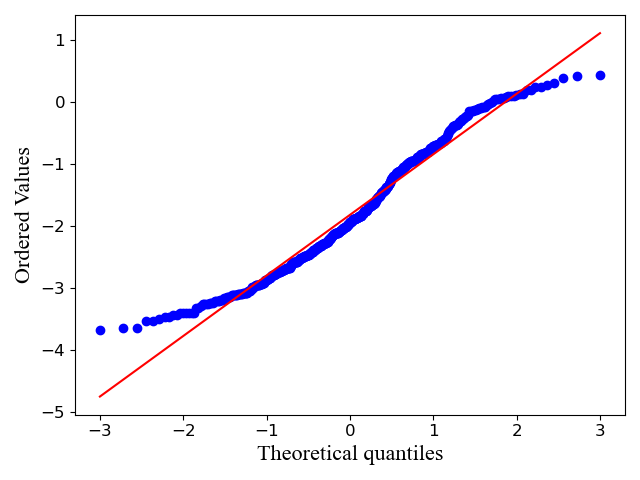}}
\subfigure[PP plot for UVW2 flux]{\label{fig:4pt2}\includegraphics[width=0.49\textwidth]{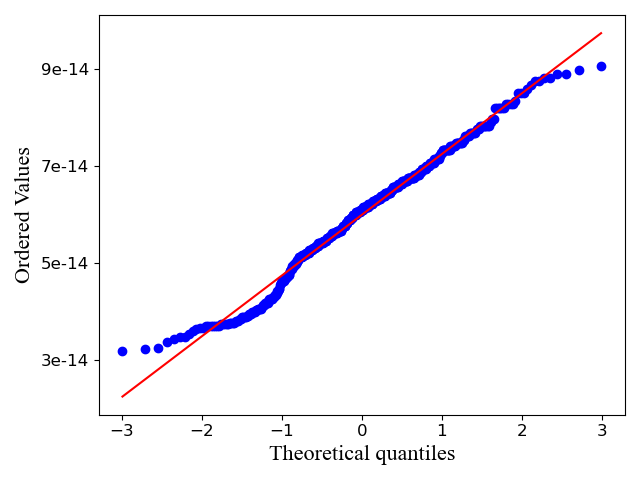}}
\subfigure[ECDF for X-ray log count rates]{\label{fig:4pt3}\includegraphics[width=0.49\textwidth]{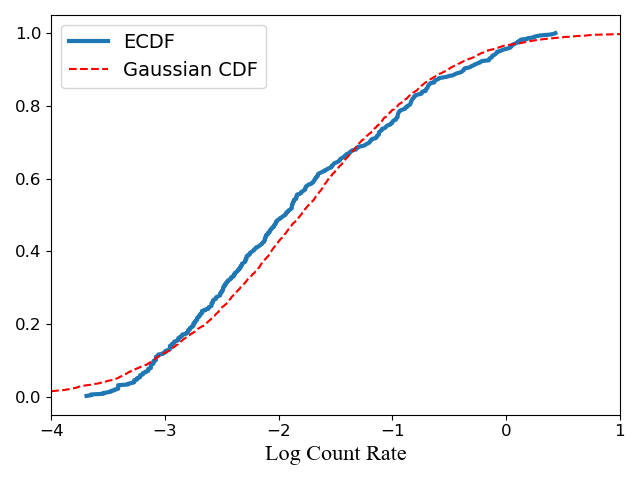}}
\subfigure[ECDF for UVW2 flux]{\label{fig:4pt4}\includegraphics[width=0.49\textwidth]{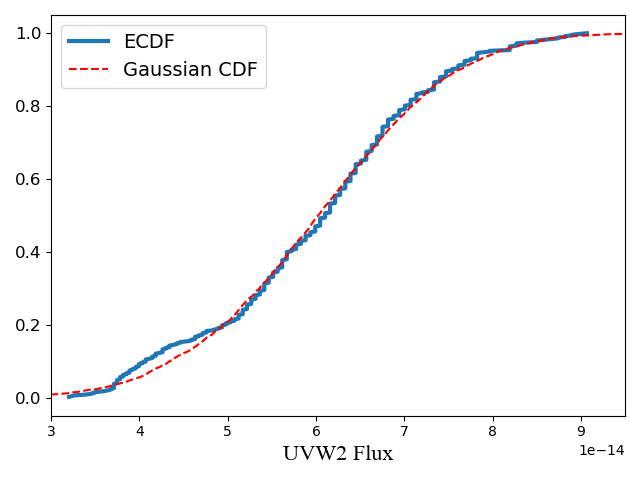}}  
\caption{P-P plots and ECDFs for X-ray log count rates and UVW2 flux. Graphical tests of Gaussianity. In the case of the P-P plots, proximity to the line is an indicator of Gaussianity. In the case of the ECDF plots, resemblance to the cumulative distribution function of a Gaussian is indicative of Gaussianity.}
\label{PP Plots}
\end{figure}

In \autoref{PP Plots} we show probability-probability (P-P) plots and empirical cumulative distributions functions (ECDFs) as graphical distribution tests for Gaussianity. It may be observed qualitatively that both X-ray band log count rates and UVW2 flux are well-modelled by a Gaussian distribution.

\newpage

\section{Kernels}
\label{kern_rat}

\begin{figure}[h!]
\centering
\subfigure[Kernel autocorrelation functions]{\label{fig:4k}\includegraphics[width=0.4\textwidth]{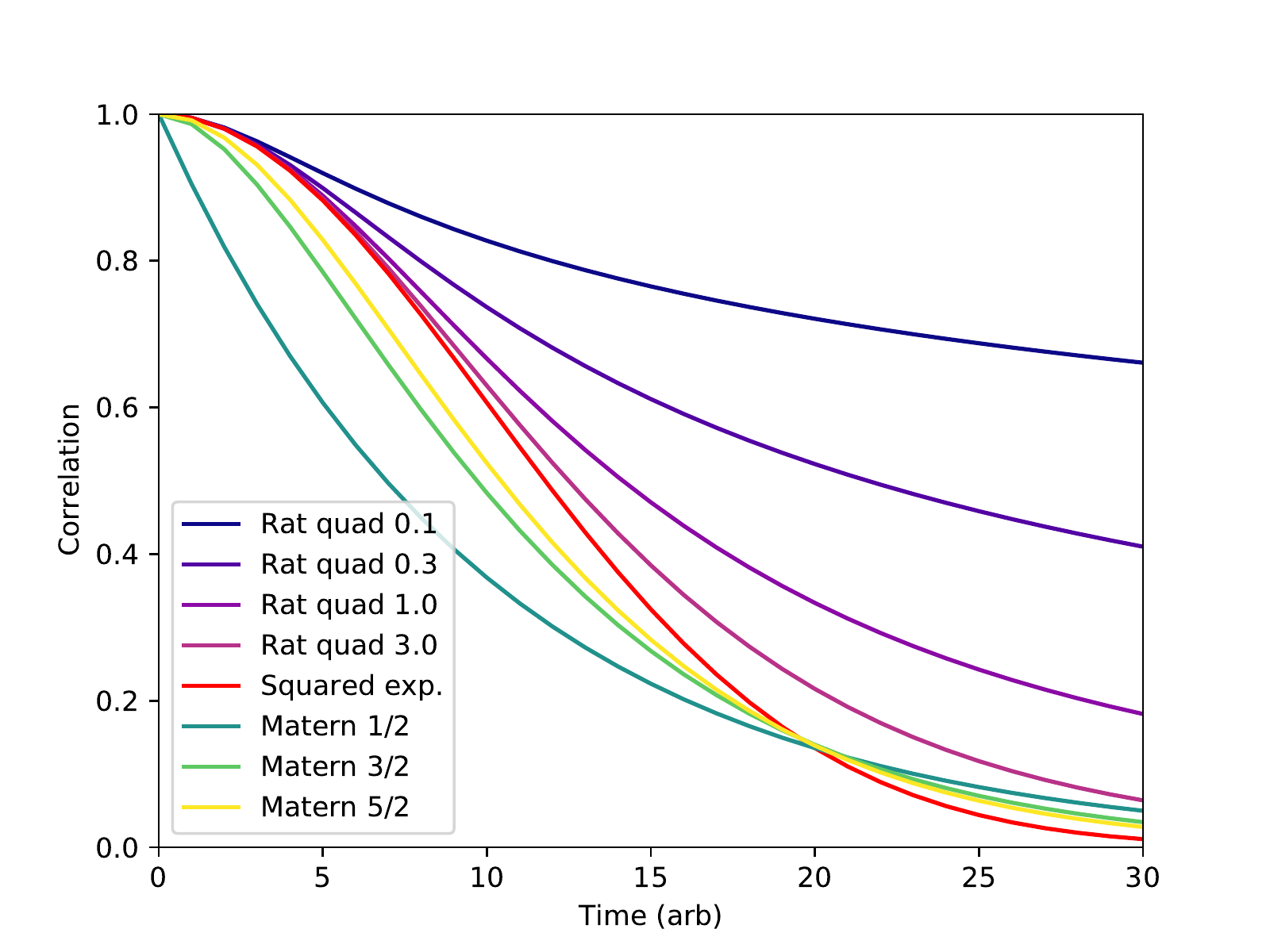}}
\subfigure[Kernel log autocorrelation functions]{\label{fig:3k}\includegraphics[width=0.4\textwidth]{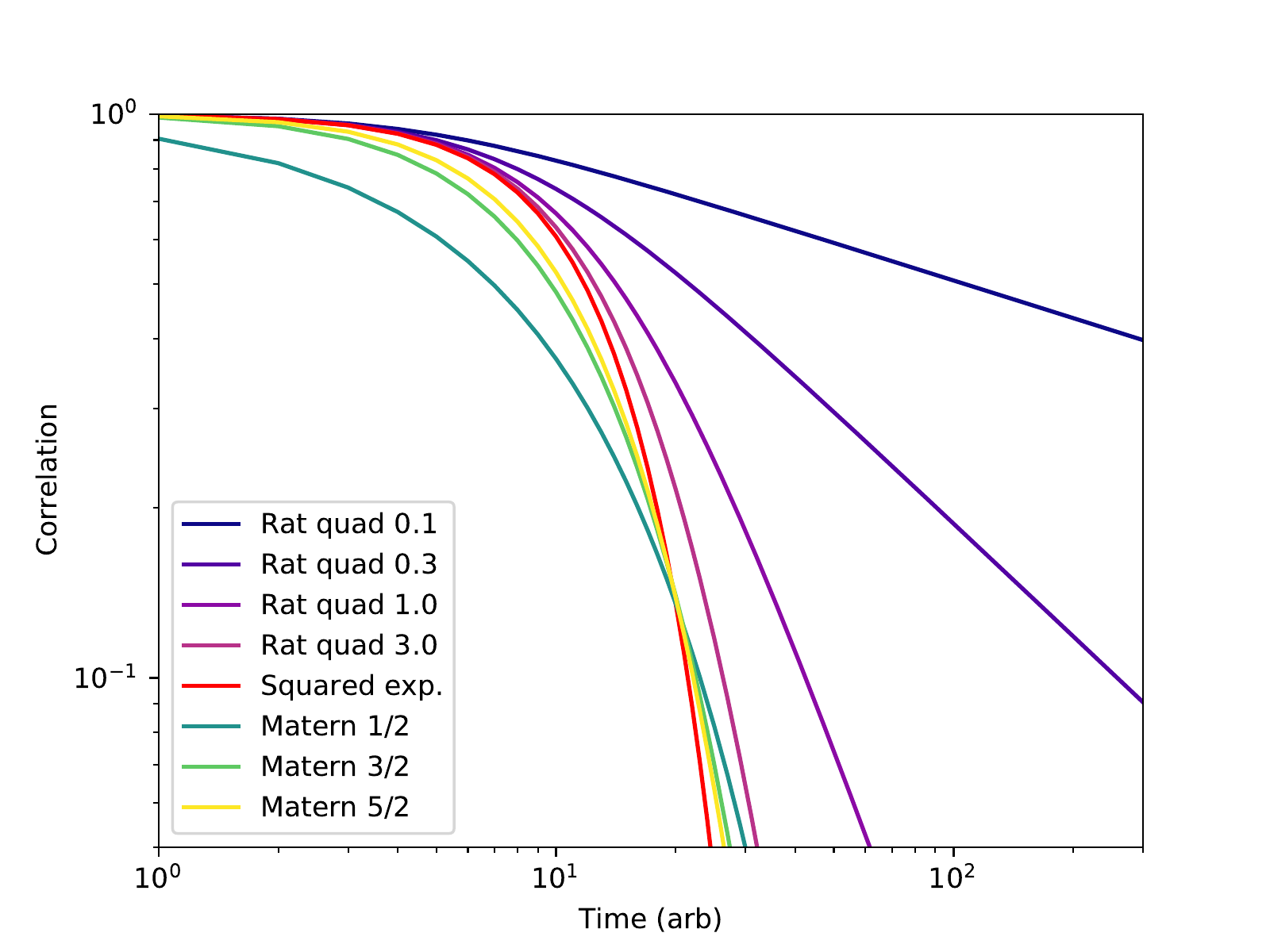}}
\subfigure[Kernel PSDs]{\label{fig:5k}\includegraphics[width=0.4\textwidth]{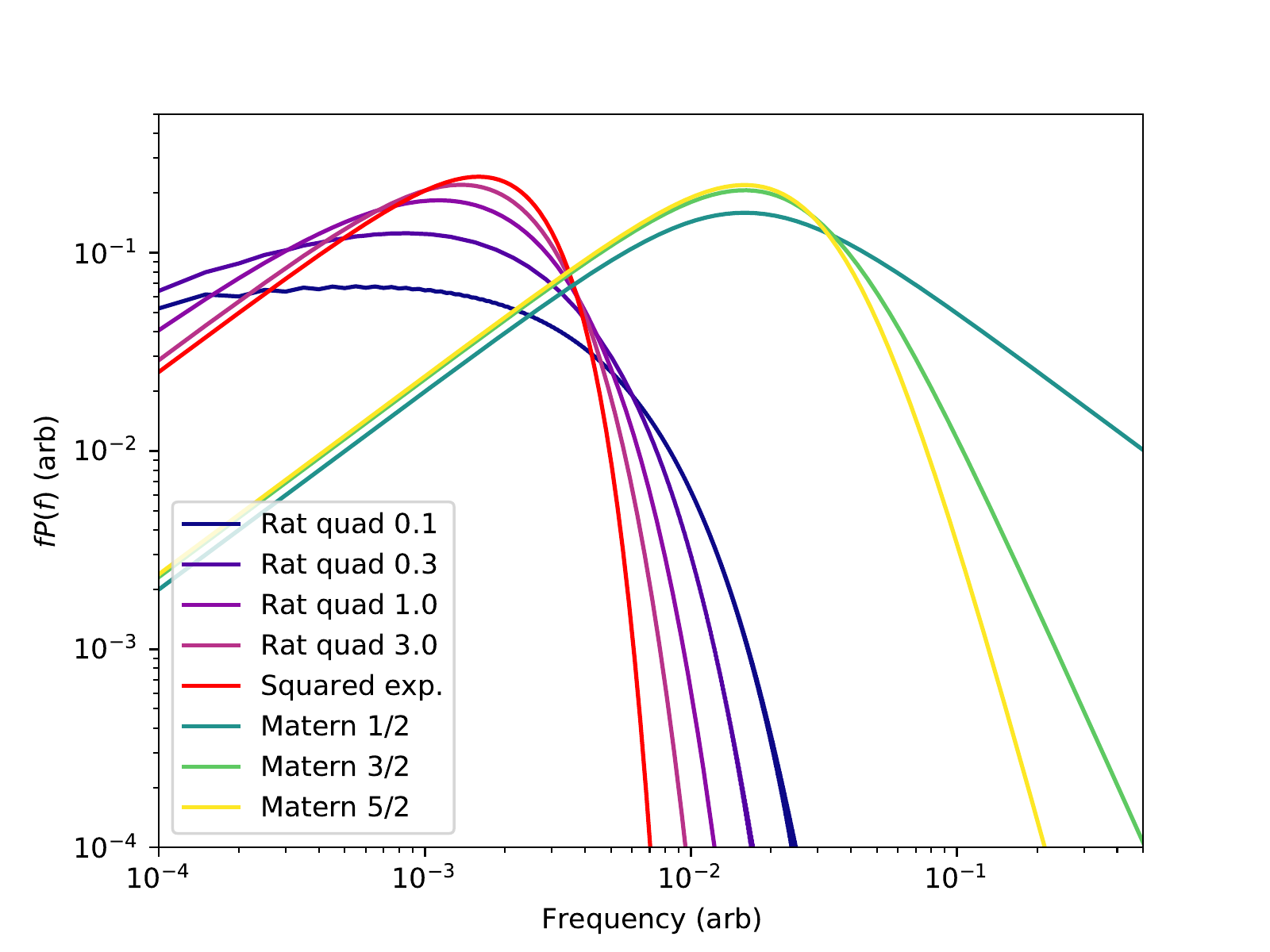}}
\caption{Kernel autocorrelation functions and PSDs. The rational quadratic kernel is plotted for different values of the $\alpha$ parameter. The Matern kernel plots in the PSD figure are offset by a factor of 10 for clarity. A PSD of $f^{-2}$ will match the high frequency part of the Matern $\frac{1}{2}$ kernel and the rational quadratic is endowed with additional flexibility to model PSDs by virtue of its $\alpha$ parameter. Such characteristics may explain why these kernels are preferred in the simulation study.}
\label{kern_acfs}
\end{figure}

\end{document}